\documentclass[%
 reprint,
 amsmath,amssymb,
 aps,
]{revtex4-2}

\usepackage{graphicx}
\usepackage{dcolumn}
\usepackage{bm}
\usepackage{xcolor}
\usepackage{subfigure}
\usepackage{amsfonts,amsmath,amssymb}
\usepackage{comment}
\usepackage{mathrsfs}
\usepackage{adjustbox}

\newcommand{\phiNNa}{\phi_{\theta_1}}

\newcommand{\tphiNNb}{\tilde{\phi}_{\theta_2}}
\newcommand{\phiNNc}{\phi_{\theta}}

\newcommand{\bfr}{\mathbf{r}}

\usepackage{booktabs}

\begin{document}

\preprint{APS/123-QED}


\title{
Exploring the Nexus of Many-Body Theories through Neural Network Techniques: the Tangent Model
}

\author{Senwei Liang}
\email{SenweiLiang@lbl.gov}
\affiliation{%
 Applied Mathematics and Computational Research Division\\
 Lawrence Berkeley National Laboratory 
}%

\author{Karol Kowalski}%
\email{karol.kowalski@pnnl.gov}
\affiliation{%
Physical Sciences Division,\ Pacific Northwest National Laboratory, Richland, Washington 99354, USA 
}%


\author{Chao Yang}
\email{CYang@lbl.gov}
\affiliation{%
 Applied Mathematics\ and Computational Research Division\\
 Lawrence Berkeley National Laboratory 
}%

\author{Nicholas P. Bauman}
\email{nicholas.bauman@pnnl.gov}
\affiliation{%
Physical Sciences Division, Pacific Northwest National Laboratory, Richland, Washington 99354, USA 
}%

%


\date{\today}

\begin{abstract}
In this paper, we present a physically informed neural network representation of the effective interactions associated with coupled-cluster downfolding models to describe chemical systems and processes.  The neural network representation not only allows us to evaluate the effective interactions efficiently for various geometrical configurations of chemical systems corresponding to various levels of complexity of the underlying wave functions, but also reveals that the bare and effective interactions are related by a tangent function of some latent variables. We refer to this characterization of the effective interaction as a tangent model.  We discuss the connection between this tangent model for the effective interaction with the previously developed theoretical analysis that examines the difference between the bare and effective Hamiltonians in the corresponding active spaces.

\end{abstract}

\maketitle

\section{Introduction}

Characterizing effective interactions within low-dimensional subspaces as part of various downfolding and embedding algorithms has become a key component in elucidating the intricate chemical and physical processes occurring across different spatial and temporal scales. These models aim to accurately account for correlation effects in composite systems by formulating effective Hamiltonians that incorporate dressed interactions that encompass long- and short-range effects, screening, and collective many-body phenomena.  A significant effort has been achieved in various downfolding \cite{shinaoka2015accuracy,aryasetiawan2022downfolding,zheng2018real,chang2024downfolding,romanova2023dynamical} and embedding methodologies \cite{georges1996dynamical,kotliar2006electronic,gull2011continuous,zgid2011dynamical,lan2015communication,knizia2012density,knizia2013density} (see also Refs.~\cite{libisch2014embedded,wesolowski2015frozen,govind1998accurate,fornace2015embedded,hegely2016exact}).


The recent advancements in coupled-cluster theory and the development of CC-driven downfolding methods have created new opportunities for defining effective/downfolded Hamiltonians \cite{kowalski2018properties,bauman2019downfolding,kowalski2020sub,kowalski2021dimensionality,bauman2022coupled,kowalski2023quantum}, similar to hierarchical classes of CC approximations used to calculate ground-state energies. Recent research has shown that the effective Hamiltonian formalism is naturally encoded in standard single-reference CC methods
\cite{coester58_421,coester60_477,cizek66_4256,paldus1972correlation,purvis82_1910,raghavachari89_479,paldus07,crawford2000introduction,bartlett07_291}
or unitary CC extensions based on the double unitary coupled-cluster (DUCC) Ansatz \cite{bauman2019downfolding,kowalski2020sub}. The latter formulations result in the Hermitian form of the effective Hamiltonians, which have been utilized in early demonstrations of quantum computing simulations to provide a more realistic description of chemical problems that require basis set sizes beyond the capabilities of brute force quantum approaches requiring approximately O($N_{\rm orb}$) number of qubits, where $N_{\rm orb}$ designates the number of correlated orbitals.
The non-Hermitian CC downfolding approach \cite{kowalski2018properties,kowalski2021dimensionality,kowalski2023sub} has recently been extended to static-type embedding procedures \cite{peng2024integrating,shee2024static}.
Another critical application area involves using deep neural networks to predict the many-body form of the Hermitian downfolded Hamiltonian. In this process, the neural network is trained using tensors that define the bare (input) and downfolded (output) representations of the Hamiltonians in small dimensionality active spaces.

In recent studies \cite{liang2024effective}, we have demonstrated the potential of neural network (NN) techniques in replacing the computationally expensive process of calculating the many-body form of the downfolded Hamiltonian. The cost of this process is typically proportional to the polynomial $(N_{\rm orb})^K$, where $N_{\rm orb}$ represents the number of orbitals and  $K\ge 6$ for accurate simulations. Our approach involves training a customized NN, which we call VNet, using bare and downfolded interactions within the active space defined by $N_{\rm orb}^{\rm (act)}$ orbitals. We have shown that a small amount of data, corresponding to a limited number of geometrical configurations of small molecules, is adequate for training VNet to interpolate and extrapolate downfolded Hamiltonians to other molecular geometries. The input/output parameter space for two-body effective interactions  in this approach scales proportionally to 
$N_{\rm orb}^{\rm (act)}\times \ell +   \ell \times \ell$ (where $\ell$ is the parameter space for approximating active molecular orbitals using neural networks; additionally, kernels defining effective two-body interactions are represented by $\ell \times \ell$ matrices).
Furthermore, NN methods have enabled us to identify functional relationships between bare and downfolded two-body interactions based on the differences in eigenvalues of their corresponding four-dimensional tensors (re-represented as two-dimensional kernels). Based on these observations, we have postulated a so-called tangent model (\texttt{tan}-model) to 
characterize the nature of the dressed effective interaction relative to the bare interaction,
a novel approach that has not been explored previously.  In this study, we will assess the performance of the \texttt{tan}-model in its application to basic benchmark systems. 

A crucial part of our analysis involves establishing connections between the \texttt{tan}-model and multi-reference formalisms utilized in nuclear theory and quantum chemistry applications. 
The VNet and related NN techniques are important because they allow for the possibility of finding confluence between various many-body models to construct effective Hamiltonian representations originating in distinct applications of quantum mechanics.
The proposed framework presents new opportunities for re-purposing single-reference coupled-cluster theory and the associated computational tools, such as symbolic many-body algebra systems (SymGen of Ref.~\cite{bylaska2024electronic} and available at \url{https://github.com/npbauman/SymGen}) and parallel tensor contraction libraries (Tensor Algebra for Many-body Methods (TAMM) of Ref.~\cite{mutlu2023tamm}), to simulate effective interactions in neural network environments.

\section{Theory}

In this section, we will briefly discuss the CC downfolding theory in its Hermitian formulations based on double unitary CC Ansatz (DUCC) \cite{bauman2019downfolding,kowalski2020sub,kowalski2021dimensionality,bauman2022coupled,kowalski2023quantum} 
where 
\begin{equation}
|\Psi\rangle = e^{\sigma_{\rm ext}(\mathfrak{h})}  e^{\sigma_{\rm int}(\mathfrak{h})} |\Phi\rangle \;,
\label{eq8}
\end{equation}
where  ($\sigma_{\rm int}(\mathfrak{h})$) and ($\sigma_{\rm ext}(\mathfrak{h})$)  refer to anti-Hermitian cluster operators. The index $\mathfrak{h}$ refers to excitation sub-algebras \cite{kowalski2018properties} generating all excited configurations in the complete active space (CAS) when acting on the reference determinant $|\Phi\rangle$.
The partitioning of the cluster operator into internal
and external parts has been initially introduced in the
context of the active-space (or state-selective) CC approaches  in Refs.~\cite{oliphant1992implementation,oliphant1993multireference,pnl93,adamowicz1998state} and was invoked as a selection mechanism for high-rank
cluster amplitudes. This partitioning provides a starting point for deriving the non-Hermitian variant of the CC downfolding. 
As discussed in Ref.~\cite{kowalski2020sub}, the DUCC Ansatz allows one to build (once certain conditions are met) an effective Hamiltonian in the $\mathfrak{h}$-generated CAS
\begin{equation}
    H^{\rm D}(\mathfrak{h}) = (P+Q_{\rm int}(\mathfrak{h}))
    e^{-\sigma_{\rm ext}(\mathfrak{h})} H
    e^{\sigma_{\rm ext}(\mathfrak{h})}
    (P+Q_{\rm int}(\mathfrak{h}))
    \label{eq9}
\end{equation}
with the lowest eigenvalue being exact energy $E$ of the system once exact form of $\sigma_{\rm ext}(\mathfrak{h})$ is known. 
The $P+Q_{\rm int}(\mathfrak{h})$ is a projection operator onto $\mathfrak{h}$-generated CAS 
($P=|\Phi\rangle\langle\Phi|$ and $Q_{\rm int}(\mathfrak{h})$ represents a projection operator onto all excited configurations with respect to $|\Phi\rangle$ in the $\mathfrak{h}$-generated CAS).
The above form of downfolded/effective Hamiltonian acting in the active space  can also be employed to evaluate approximations to exact energy when approximate forms of 
$\sigma_{\rm ext}(\mathfrak{h})$ are available.

In contrast to the second-quantized form of the bare Hamiltonian in the active space ($H^{\rm B}(\mathfrak{h})$), 
\begin{widetext}
\begin{equation}
H^{\rm B}(\mathfrak{h}) = \Gamma^{\rm B}_0(\mathfrak{h})+ \sum_{PQ} h^P_Q(\mathfrak{h}) a_P^{\dagger} a_Q + \frac{1}{4} \sum_{P,Q,R,S} v^{PQ}_{RS}(\mathfrak{h}) a_P^{\dagger} a_Q^{\dagger} a_S a_R \;,
\label{eq10b}
\end{equation}
\end{widetext}
the second-quantized representation of $H^{\rm D}(\mathfrak{h})$
\begin{widetext}
\begin{equation}
H^{\rm D}(\mathfrak{h})= \Gamma^{\rm D}_0(\mathfrak{h}) +
\sum_{PQ} g^P_Q(\mathfrak{h}) a_P^{\dagger} a_Q + \frac{1}{4} \sum_{P,Q,R,S} k^{PQ}_{RS}(\mathfrak{h}) a_P^{\dagger} a_Q^{\dagger} a_S a_R 
+ \frac{1}{36} \sum_{P,Q,R,S,T,U}
l^{PQR}_{STU}(\mathfrak{h}) a_P^{\dagger} a_Q^{\dagger} a_R^{\dagger} a_U a_T a_S + \ldots \;.
\label{eq10}
\end{equation}
\end{widetext}
contains higher-than-pairwise interactions. 
In the above formulas, $P$, $Q$, $R$, $S$, $\ldots$ refer to active spin-orbitals and 
$a_P^{\dagger}$ ($a_P$) represents the creation (annihilation) operators for an electron in $P$-th active spin-orbital.
For many practical applications, the inclusion of one- and two-body interaction is sufficient to capture dynamical correlation effects that involve collective effects outside of active space. 

Recently, we explored the utilization of NN techniques (henceforth referred to as VNet) to by-pass numerically intensive process of evaluating $H^{\rm D}(\mathfrak{h})$  using $H^{\rm B}(\mathfrak{h})$ as an input  \cite{liang2024effective}, i.e.,
\begin{equation}
H^{\rm B}(\mathfrak{h})
\xrightarrow{\text{VNet}}
H^{\rm D}(\mathfrak{h}) \;.
\label{eq13}
\end{equation}
To this end, we focused on pairwise interactions 
\begin{equation}
\lbrace v^{PQ}_{RS}(\mathfrak{h}) \rbrace
\xrightarrow{\text{VNet}} 
 \lbrace  k^{PQ}_{RS}(\mathfrak{h})  \rbrace 
 \;.
\label{eq13b}
\end{equation}
and used a subset of pairwise interactions corresponding to several molecular geometries symbolically denoted as $\lbrace {\bf R}_i \rbrace_{i=1}^{M_T}$, i.e., 
\begin{equation}
\lbrace v^{PQ}_{RS}(\mathfrak{h},{\bf R}_i) \rbrace_{i=1}^{M_T}
\xrightarrow{\text{VNet training}} 
 \lbrace  k^{PQ}_{RS}(\mathfrak{h},{\bf R}_i)  \rbrace_{i=1}^{M_T} 
 \;.
\label{eq13c}
\end{equation}
to train VNet.
To reduce the number of parameters needed to express one and two-body interactions
$h^P_Q$/$v^{PQ}_{RS}$ and $g^P_Q$/$k^{PQ}_{RS}$
we used the Mulliken type notation for the orbital representation of one- and two-body interactions, namely $(p|q)_B$/$(pq|rs)_B$ and
$(p|q)_D$/$(pq|rs)_D$. For notational simplicity, we will also skip the $\mathfrak{h}$ index.
In this paper, we will extend our analysis to one-body interactions, i.e.
\begin{widetext}
\begin{equation}
\lbrace (p|q)_B, (pq|rs)_B \rbrace
\xrightarrow{\text{VNet}} 
 \lbrace (p|q)_D, (pq|rs)_D  \rbrace
 \;.
\label{eq13c}
\end{equation}
\end{widetext}
Tensors defining bare interactions, $(p|q)_B$ and $(pq|rs)_B$, are defined as follows:
\begin{eqnarray}
&(p|q)_B &= \int d\bfr  \phi_p(\bfr) h(\bfr) \phi_q(\bfr) \;, \label{eq:hpq} \\
&(pq|rs)_B
&= \int d\bfr \int d\bfr'  \phi_p(\bfr)\phi_q(\bfr) \frac{1}{|\bfr-\bfr'|} \phi_r(\bfr') \phi_s(\bfr'),
\label{eq:vpqrs}
\end{eqnarray}
where the one-body Hamiltonian $h(\bfr)$ and Coulomb interactions $\frac{1}{|\bfr-\bfr'|}$ ar the integration kernels. We will utilize Eqs.~(\ref{eq:hpq})-(\ref{eq:vpqrs}) to build the representation of bare and effective interactions for NN models.

\section{Neural Network representations of  one- and two-body interactions}

Following the discussion of Ref.~\cite{liang2024effective}, the proposed NN-based VNet algorithm employs the following representation for the tensors given by Eqs.~(\ref{eq:hpq})-(\ref{eq:vpqrs}). For example, in the case of tensors that define the bare two-body interaction in the orbital representation, we express them as:
\begin{eqnarray}
(pq|rs)_B
&=& \left[\phi_p\odot\phi_q\right]^T W^B \left[ \phi_r \odot \phi_s\right] \;,
\label{nnv1}
\end{eqnarray} 
where  $\phi_p$, $\phi_q$, $\phi_r$, $\phi_s$ are vectors (or pseudo-orbitals) of length $\ell$, chosen from a set of $N_{\rm orb}^{\rm (act)}$ vectors $\{\phi_1,...,\phi_{N_{\rm orb}^{\rm (act)}}\}$ 
($N_{\rm orb}^{\rm (act)}$ stands for the total number of active orbitals), 
$W^B$ is  $\ell \times \ell$ symmetric matrix representing computational kernels for bare interaction, and $\phi_p\odot\phi_q$ denotes the element-wise product of $\phi_p$ and $\phi_q$. 

For tensors representing the effective two-body interaction, we employ the following expression:
\begin{eqnarray}
\begin{split}
    (pq|rs)_D
=&& \frac{1}{2}\left([\phi_p\odot\tilde{\phi}_q]^T W^D [ \phi_r \odot \tilde{\phi}_s] \right.
\\&&+\left. [\phi_q\odot\tilde{\phi}_p]^T W^D [ \phi_s \odot \tilde{\phi}_r] \right) \;.
\end{split}
\label{nnv2}
\end{eqnarray}
Here, $\tilde{\phi}_p$ is also a vector (or pseudo-orbital) of length $\ell$ that is closely related to $\phi_p$, but not identical to it. This distinction is necessary to preserve the 4-fold symmetry of the effective interaction tensors. The $W^D$ in \eqref{nnv2} is a symmetric matrix that represents the the effective interaction kernel. For simplicity, it is assumed that $W^D$ is independent of the molecular geometry. In future research, we will incorporate explicit dependencies on the molecular geometry for the $W^D$ kernel.

The procedure we use to train the VNet representation of the two-body interactions consists of two steps:
\begin{enumerate}
\item  All $\phi'$s are represented by an NN parameterized by a set of weights and biases contained in a vector $\theta$. The NN takes the orbital index as one of its inputs, encoded as a one-hot vector. Because molecular orbitals depend on the geometry of the molecule, the NN should also take the geometry represented by a parameter $\mathbf{R}$ as input. 
 Since the $W$ kernels are assumed geometry independent, the $\phi$ vectors encapsulate the geometry dependence needed to reproduce one- and two-body interactions for molecular geometries of interest. We will denote this NN by $\phi_{\theta}(i,\mathbf{R})$, where $i$ represents an orbital index. The output of $\phi_{\theta}(i,\mathbf{R})$ contains $\ell$ real numbers.  The architecture of this NN is shown in Figure~\ref{fig:nnstructure}(a).  


We use the bare interaction tensors for several different geometries $\{\mathbf{R}\}$ to train the NN to obtain an initial set of parameters $\theta$ and the $W^B$ matrix, as shown in Figure~\ref{fig:nnstructure}(b). For each geometry, the training data consists of elements of 
$(pq|rs)_B$
that belong to one nonsymmetric unit. Note that the algebraic decomposition of the bare two-body interaction tensor in \eqref{nnv1} eliminates the physical spatial coordinate $\bfr$ of a molecular orbital and replaces it with a latent spatial coordinate.

\item 
In the second step, we use the effective two-body interaction tensor elements 
$(pq|rs)_D(\mathbf{R})$
associated with a few selected geometries $\mathbf{R}_i$, $i=1,2,...,n_{\rm ref}$, to obtain the two NNs ($\phi_{\theta_1}(i,\mathbf{R})$ and $\tilde{\phi}_{\theta_2}(i,\mathbf{R})$) and to construct an approximation of the effective interaction kernel $W^D$, as shown in Figure~\ref{fig:nnstructure}(c). The NNs $\phi_{\theta_1}$ and $\tilde{\phi}_{\theta_2}$ are used to represent $\phi_i$ and $\tilde{\phi}_i$ in \eqref{nnv2}, respectively. Notably, both $\theta_1$ and $\theta_2$ are initialized with the parameters $\theta$ obtained from step 1, and $W^D$ is initialized with $W^B$, which ensures that the initial model resembles the bare tensor. These parameters are finetuned by solving the optimization problem:
\begin{widetext}
\begin{eqnarray}
\begin{split}
\min_{W^D, \theta_1, \theta_2}  \sum_{i=1}^{n_{\rm ref}} \sum_{\{p,q,r,s\}\in \mathcal{S}} \biggl | 
(pq|rs)_D(\mathbf{R}_i)
- \frac{1}{2}\biggr(\left[\phiNNa(p,\mathbf{R}_i)\odot \tphiNNb(q,\mathbf{R}_i)\right]^T W^D \left[\phiNNa(r,\mathbf{R}_i)\odot \tphiNNb(s,\mathbf{R}_i) \right] + \\
 \left[\phiNNa(q,\mathbf{R}_i)\odot \tphiNNb(p,\mathbf{R}_i)\right]^T W^D \left[\phiNNa(s,\mathbf{R}_i)\odot \tphiNNb(r,\mathbf{R}_i) \right]\biggr) \biggr\|^2,
 \end{split}
\end{eqnarray}
\end{widetext}
where $\mathcal{S}$ is a subset of the full index set for the $N_{\rm orb}^{\rm (act)}\times N_{\rm orb}^{\rm (act)}\times N_{\rm orb}^{\rm (act)} \times N_{\rm orb}^{\rm (act)}$ tensor 
$(pq|rs)_D(\mathbf{R})$
that belongs to a nonsymmetric subunit, and $n_{\rm ref}$ is the number of selected reference geometries.  Note that  $(pq|rs)_D$ only has a 4-fold symmetry. 
Calculating $(pq|rs)_D$, in contrast to $(pq|rs)_B$,  is associated with a significant numerical overhead.

\begin{figure*}
\includegraphics[width=1.0\textwidth
]{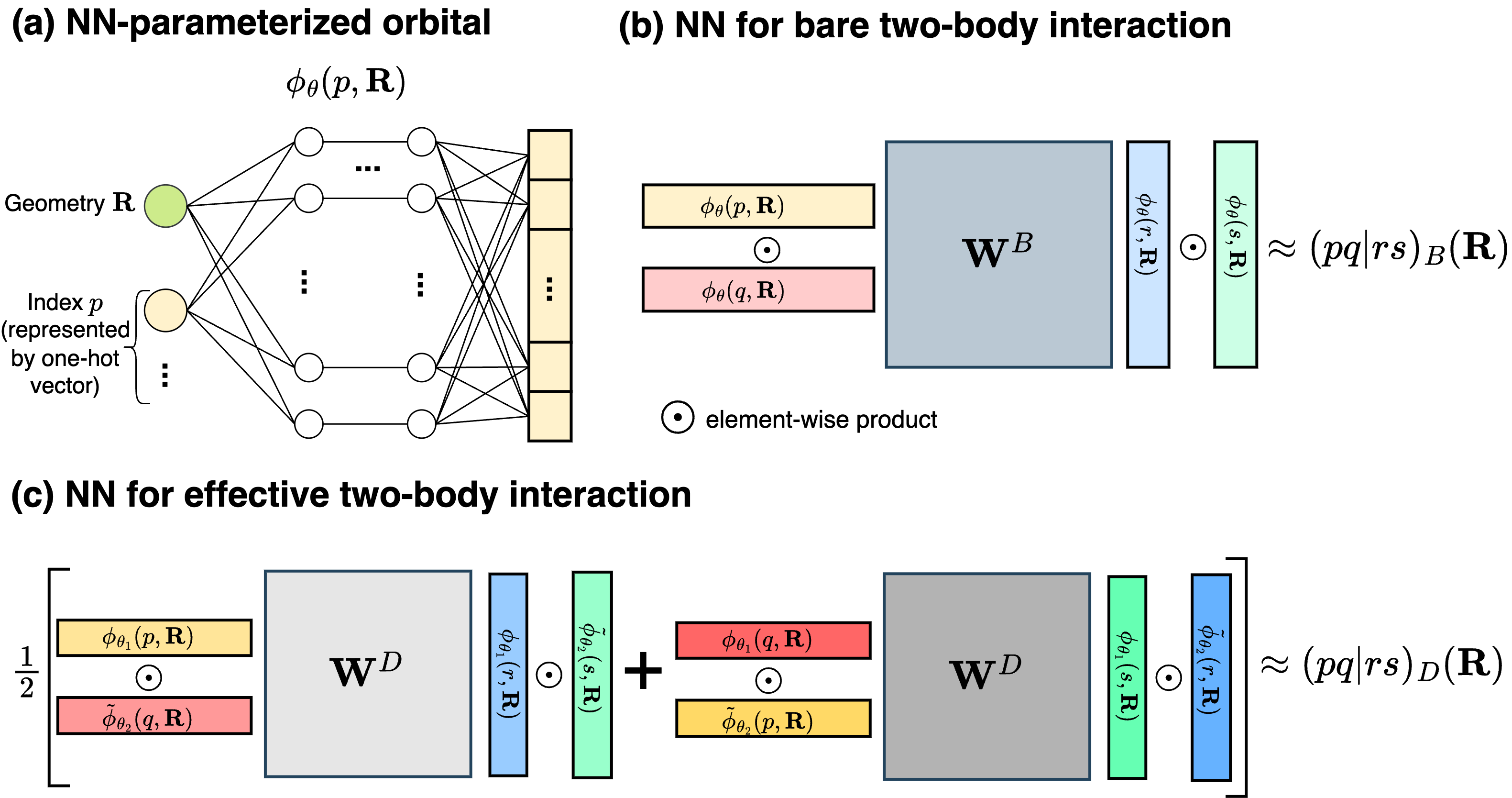}
\caption{(a) NN representation of the molecular orbital $p$. (b) The ansatz NN for approximating the bare two-body interaction tensor. (c) The ansatz NN for approximating the effective two-body interaction tensor. }
\label{fig:nnstructure}
\end{figure*}

It is important to point out that both 
$(pq|rs)_B(\mathbf{R})$
and 
$(pq|rs)_D(\mathbf{R})$
are geometry dependent. This $\mathbf{R}$ dependency translates into the dependence of $\phiNNc$, $\phiNNa$ and $\tphiNNb$ on $\mathbf{R}$.  However, both the bare and effective interaction kernels $W^B$ and $W^D$ are independent of $\mathbf{R}$. As a result, we can choose $\mathcal{S}$ to be the entire index subset for a nonsymmetric subunit of 
$(pq|rs)_D(\mathbf{R})$
and determine $W^D$ from the effective two-body interaction tensor for a few geometric configurations associated with a few different $\mathbf{R}_i$'s.  This $W^D$ can be combined with the refined NNs $\phiNNa(i,\mathbf{R}')$ and $\tphiNNb(i,\mathbf{R}')$ to predict effective two-body interaction tensors at other geometries defined by a different set of $\mathbf{R}'$.
\end{enumerate}

Let
\begin{eqnarray}
W^B &=& Z^B \Gamma^B (Z^B)^T \label{wbdec} \;, \\
W^D &=& Z^D \Gamma^D (Z^D)^T \;,
\end{eqnarray}
be the eigendecomposition of $W^B$ and $W^D$ respectively,
where the matrices $Z^B$ and $Z^D$ contain the eigenvectors of $W^B$ and $W^D$, and the diagonal matrices $\Gamma^B$ and $\Gamma^D$ contain the corresponding eigenvalues $\varepsilon_1^B,\varepsilon_2^B,...,\varepsilon_{\ell}^B$ and 
$\varepsilon_1^D,\varepsilon_2^D,...,\varepsilon_{\ell}^D$.
The numerical experiments we performed in~\cite{liang2024effective} showed that the $Z^D\simeq Z^B$ and that the eigenvalues $\varepsilon_i^D$ and $\varepsilon_i^B$ can be related using a scaled tangent function, i.e., 
\begin{equation}
 \varepsilon^D_i = \varepsilon^B_i (1+\beta \tan(\alpha (i-i_c))) \;,
 \label{tan1}
\end{equation}
for some parameters $\alpha$, $\beta$ and $i_c$. Rearranging terms yields the following expression, which suggests that the relative difference between eigenvalues of $W^D$ and $W^B$ can be described by a tangent function, i.e., 
\begin{equation}
 \varepsilon^D_i - \varepsilon^B_i = \varepsilon^B_i \beta \tan(\alpha (i-i_c)) \;.
 \label{tan2}
\end{equation}
We refer to \eqref{tan2}, which is derived from analyzing a machine-learned interaction kernel, as a \texttt{tan}-model for describing the relationship between the bare and effective interactions.

One of the purposes of this paper is to validate the \texttt{tan}-model for one-body effective interactions also. To this end, we follow steps 1 and 2 above for two-body interactions and  extend the pseudo-orbitals approach to one-body interactions using NN representation of pseudo-orbitals ($\psi_p$'s) (in general, different from pseudo-orbitals used for representing two-body interactions ($\phi_q$s)):
\begin{eqnarray}
    (p|q)_B &=& \psi_p M^B \psi_q \;, \label{nnh1} \\
    (p|q)_D &=& \psi_p M^D \psi_q \;. \label{nnh2}
\end{eqnarray}
The procedure for training the VNet representations of one-body interactions is straightforward. We use an NN, $\psi_\Theta(p, \mathbf{R})$, to represent $\psi_p$. Step 1 involves training on multiple geometries of bare one-body interaction tensors to obtain $\Theta$ and $M^B$. In step 2, $M^D$ is fine-tuned, starting from $M^B$.

We also perform an eigendecomposition of the kernel matrices $M^D$ and  $M^B$ to obtain
\begin{eqnarray}
M^B &=& X^B \Lambda^B (X^B)^T \label{wbdec} \;, \\
M^D &=& X^D \Lambda^D (X^D)^T \;,
\end{eqnarray}
where the matrices $X^B$ and $X^D$ contain the eigenvectors of $M^B$ and $M^D$, and the diagonal matrices $\Lambda^B$ and $\Lambda^D$ contain the corresponding eigenvalues $\lambda_1^B,\lambda_2^B,...,\lambda_{\ell}^B$ and 
$\lambda_1^D,\lambda_2^D,...,\lambda_{\ell}^D$. We check whether the tangent model 
\begin{equation}
 \lambda^D_i - \lambda^B_i = \lambda^B_i \delta \tan(\gamma(i-i_e)) \;,
 \label{tan21}
\end{equation}
holds for the one body interactions by optimizing the the parameter $\gamma$, $\delta$, and $i_e$ that are the ``one-body" counterparts of ``two-body" parameters 
$\alpha$, $\beta$, and $i_c$ to minimize the difference between the left and right hand sides of \eqref{tan21}.\\

\section{Numerical examples}

In the following, we validate \texttt{tan}-models for one- and two-body interactions for several benchmark systems. 
For all VNet models, we use a fully connected NN, as shown in Figure~\ref{fig:nnstructure}(a), to represent the orbital NN. This NN consists of 3 hidden layers, with each layer containing 200 neurons. The output width is $\ell=300$. SiLU activation~\cite{elfwing2018sigmoid} is used. Adam optimizer is used to train the orbital network and the interaction kernel matrix. The learning rate follows the decay of the cosine learning rate. We summarize all VNet key training configurations in Tables~\ref{tab:Onebodytraining} and \ref{tab:Twobodytraining}. 

We consider the following four benchmark systems with the cc-pVTZ basis and all electrons correlated: 
\begin{description}
    \item[H$_4$] For the linear H$_4$ molecule, the number of active orbitals is 4. We trained VNet on the bare one- (two-) body interaction tensors, covering bond lengths of nearest-neighboring hydrogen atoms from 1.8 to 3.0 a.u., in a total of 121 tensors. Then, we fine-tuned the model using the effective one- (two-) body tensors for bond lengths of 1.85, 2.25, 2.65, and 2.95 a.u.
    \item[H$_6$] For the linear H$_6$ molecule, with 6 active orbitals, we trained VNet on the bare one- (two-) body interaction tensors over bond lengths of nearest-neighboring hydrogen atoms ranging from 1.8 to 3.0 a.u., totaling 121 tensors. Subsequently, the model was fine-tuned using the effective one- (two-) body tensors at bond lengths of 1.85, 2.25, 2.65, and 2.95 a.u. 
    \item[HF] For the HF molecule with 8 active orbitals, we trained the VNet using bare two-body interaction tensors corresponding to various relative H-F bond lengths ($R_{\rm HF}/R_{\rm eq}$; $R_{\rm eq} = {\rm 1.7325 a.u.}$), spanning a range from 0.85 to 2.0. This training utilized a total of 215 different geometries. We finetuned the model using the effective interaction tensors corresponding to relative bond lengths of 0.95, 1.35, 1.65, and 1.95.
    \item[H$_2$O] For the H$_2$O molecule with 8 active orbitals, we train the VNet using bare interaction tensors corresponding to various relative O-H bond lengths for a single bond breaking. Specifically, these represent the ratio between the O-H bond length and the equilibrium O-H bond length ($R_{\rm OH}/R_{\rm eq}$; $ R_{\rm eq} = {\rm 1.84345\; a.u.}, \Theta_{\rm HOH}=110.6^{\circ}$), spanning a range from 1.1 to 2.5, covering a total of 66 different geometries. The model was then fine-tuned using effective two-body interaction tensors for relative bond lengths of 1.15, 1.45, 1.95, and 2.45.
\end{description}

\begin{figure*}
\includegraphics[width=0.45\textwidth
]{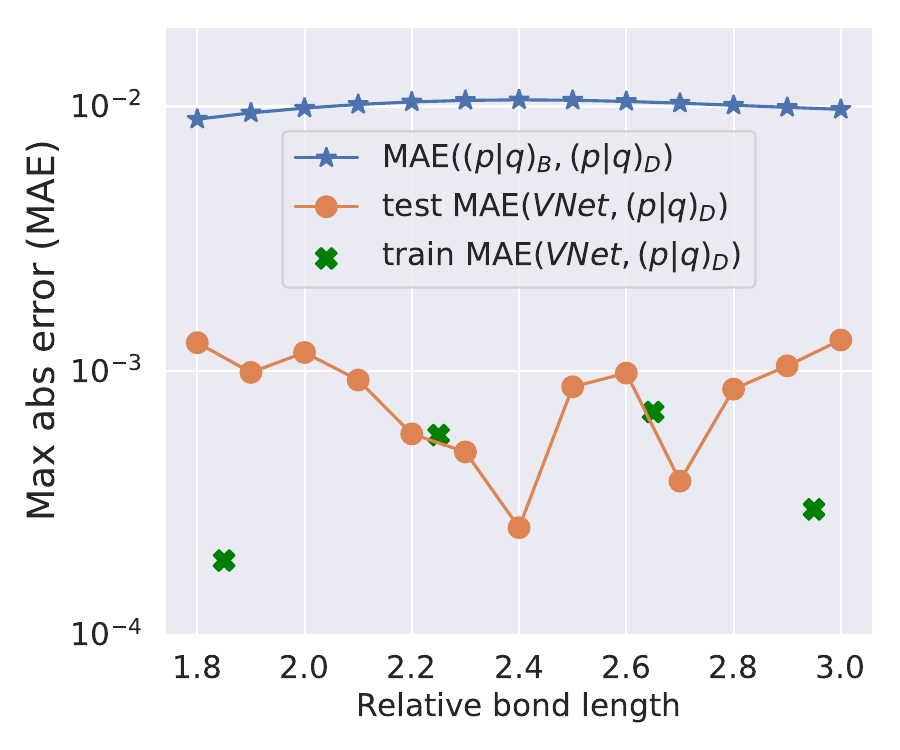}
\includegraphics[width=0.49\textwidth
]{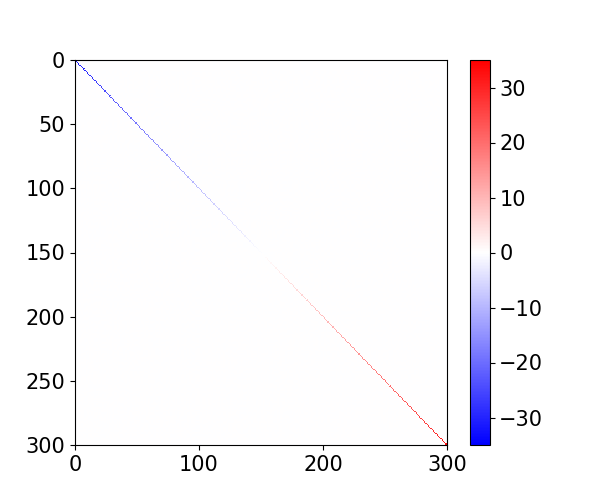}
\includegraphics[width=0.8\textwidth
]{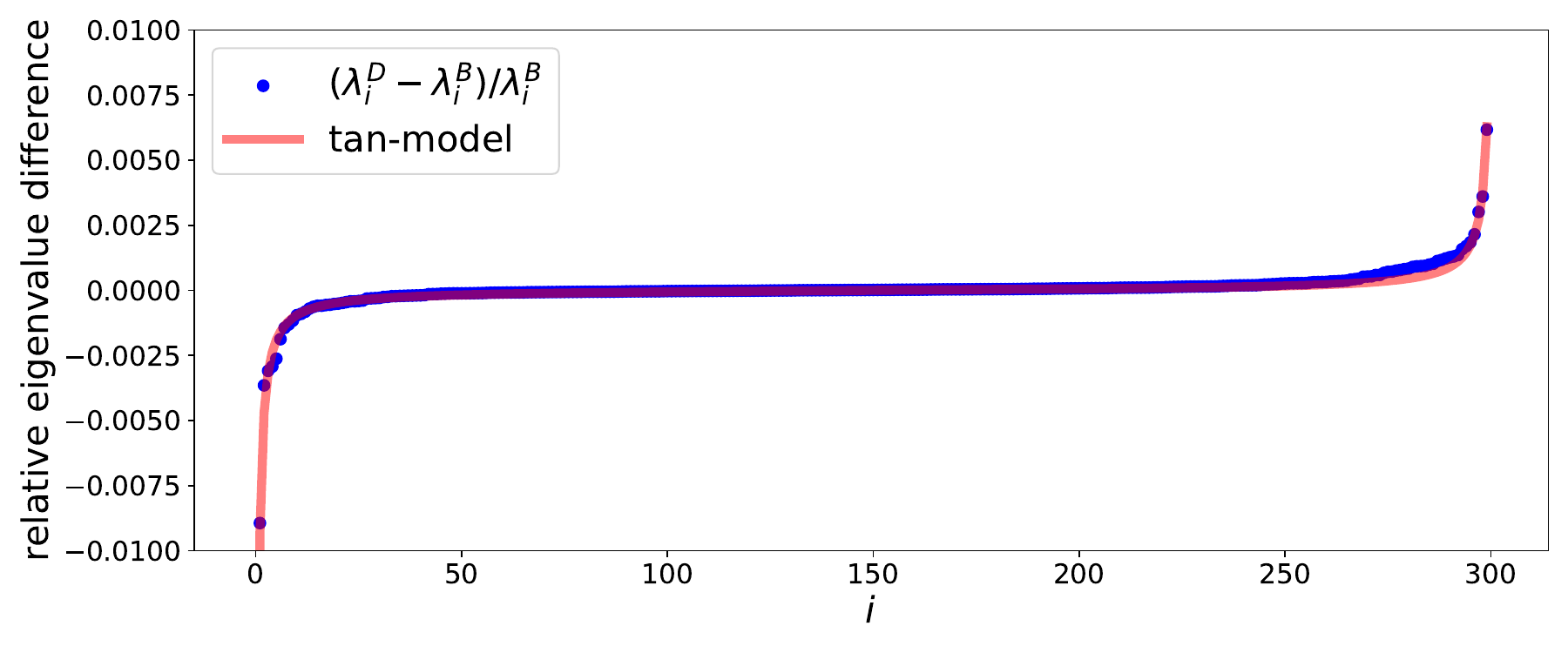}
\caption{VNet approximating the one-body interaction of the H$_4$ molecule. Top-left: Training and testing MAE of VNet across various geometries. Top-right: $(X^B)^T M^D X^B$ representation. Bottom: Empirical \texttt{tan}-model. 
}
\label{fig:h4onebody}
\end{figure*}

\begin{figure*}
\includegraphics[width=0.42\textwidth
]{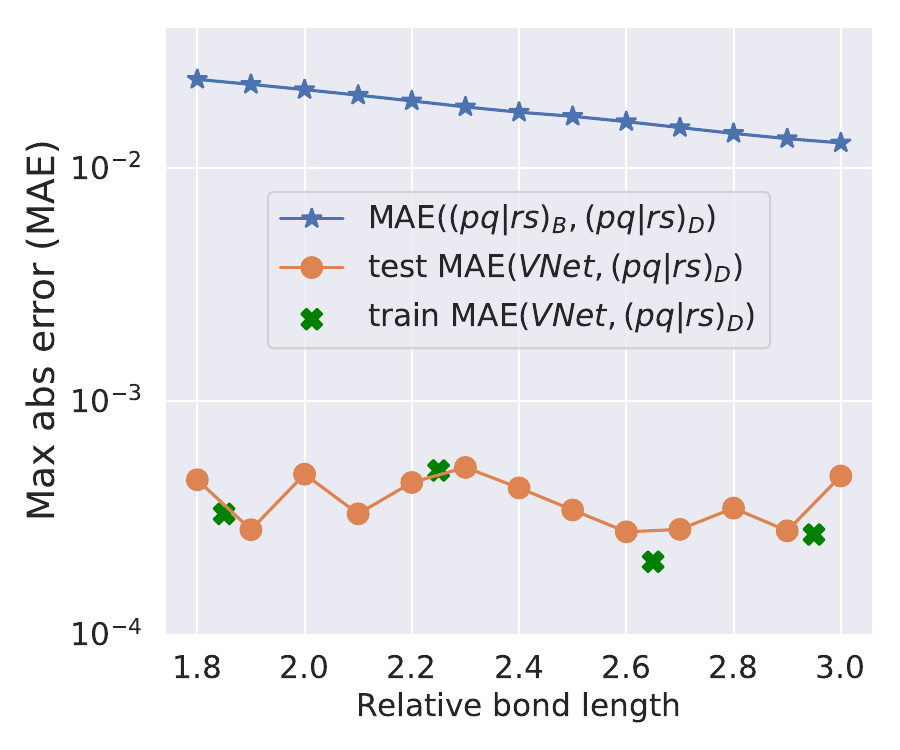}
\includegraphics[width=0.46\textwidth
]{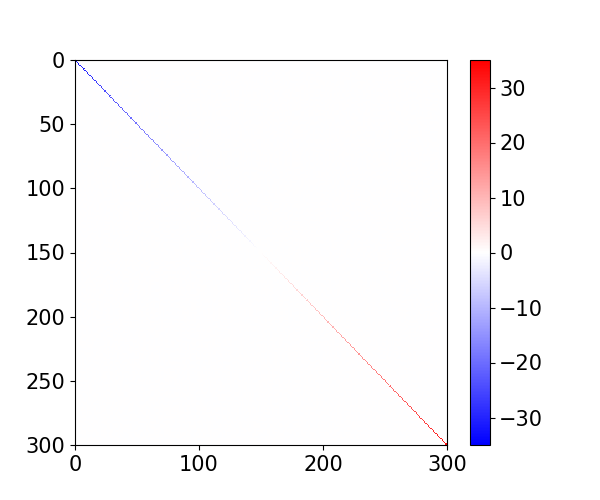}
\includegraphics[width=0.82\textwidth
]{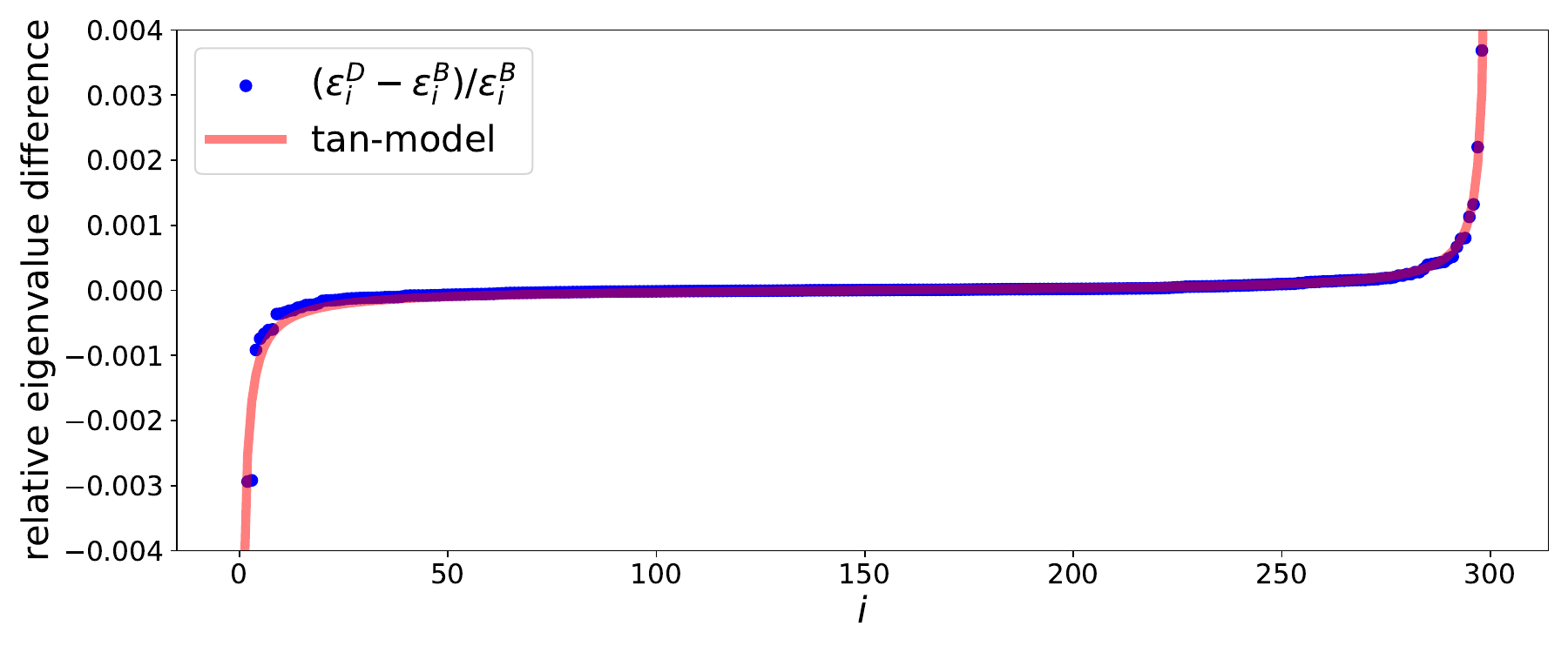}
\caption{VNet approximating the two-body interaction of the H$_4$ molecule. Top-left: Training and testing MAE of VNet across various geometries. Top-right: $(Z^B)^T W^D Z^B$ representation. Bottom: Empirical \texttt{tan}-model. 
}
\label{fig:h4twobody}
\end{figure*}

\begin{figure*}
\includegraphics[width=0.45\textwidth
]{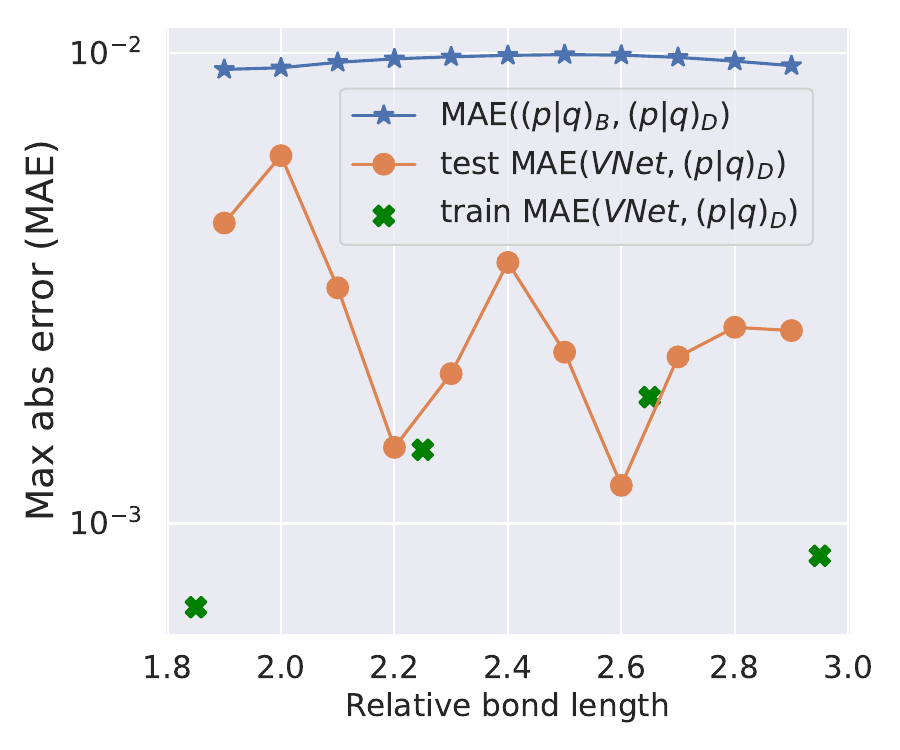}
\includegraphics[width=0.49\textwidth
]{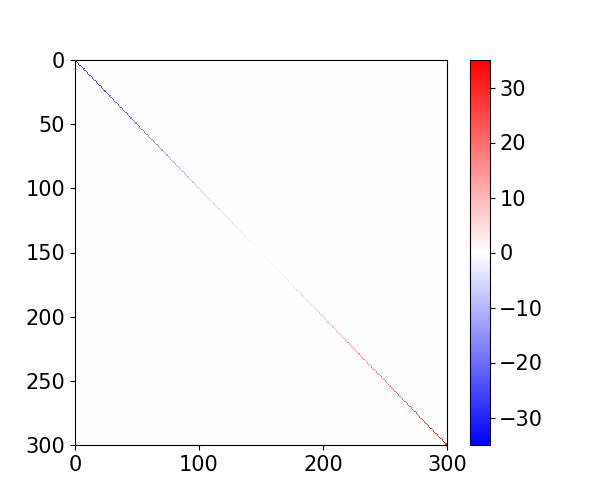}
\includegraphics[width=0.8\textwidth
]{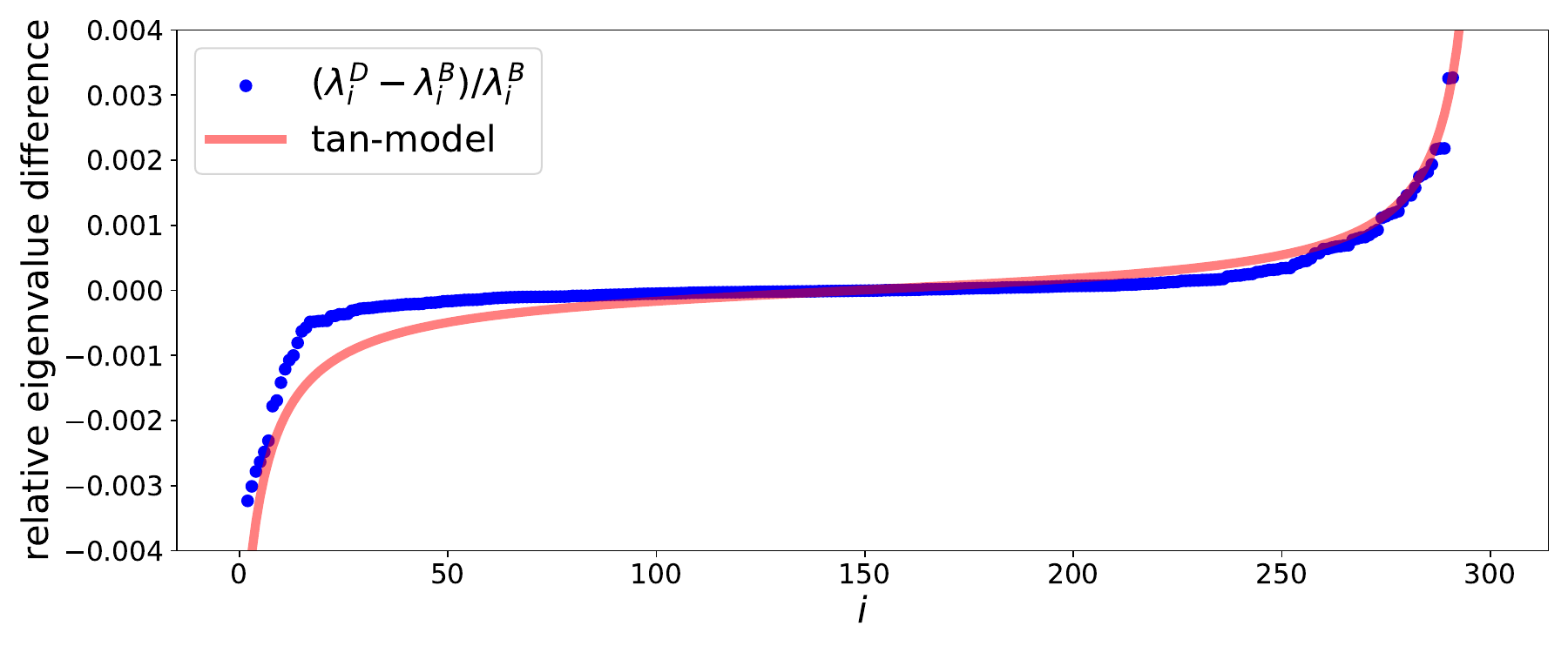}
\caption{VNet approximating the one-body interaction of the H$_6$ molecule. Top-left: Training and testing MAE of VNet across various geometries. Top-right: $(X^B)^T M^D X^B$ representation. Bottom: Empirical \texttt{tan}-model. 
}
\label{fig:h6onebody}
\end{figure*}

\begin{figure*}
\includegraphics[width=0.45\textwidth
]{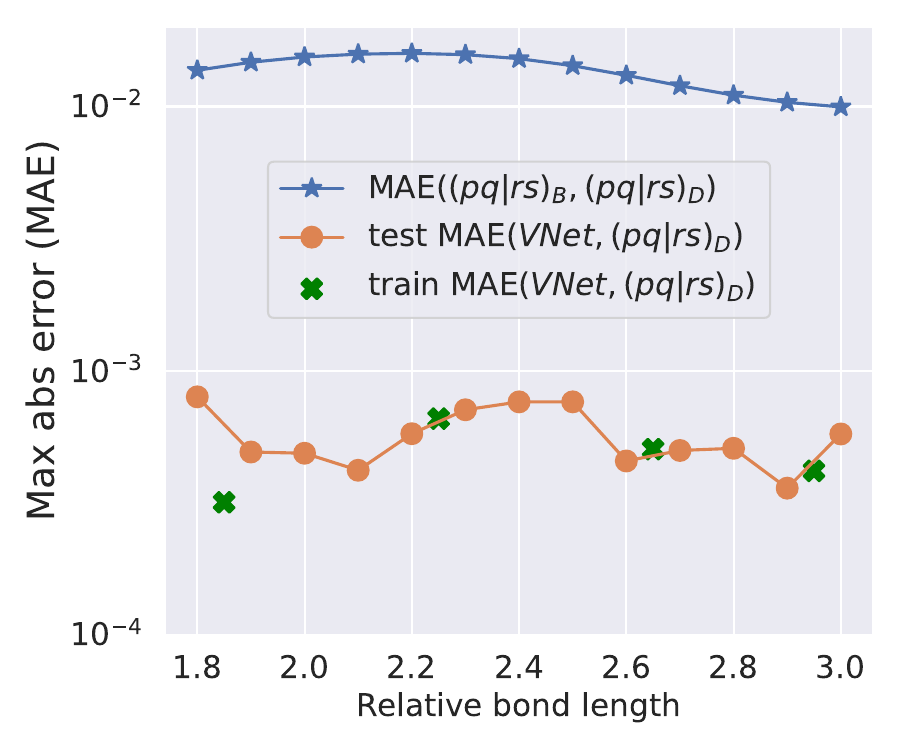}
\includegraphics[width=0.49\textwidth
]{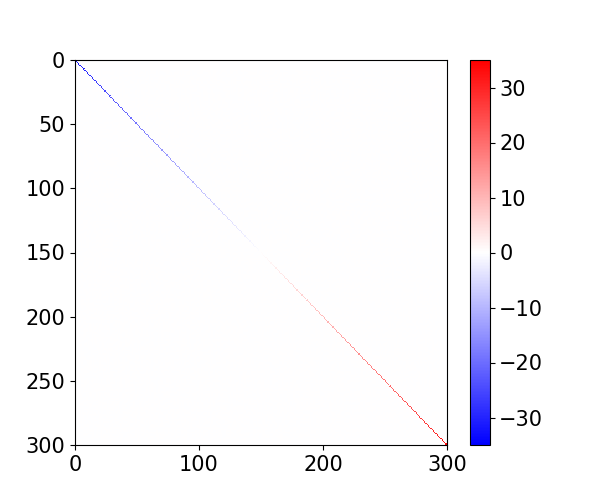}
\includegraphics[width=0.8\textwidth
]{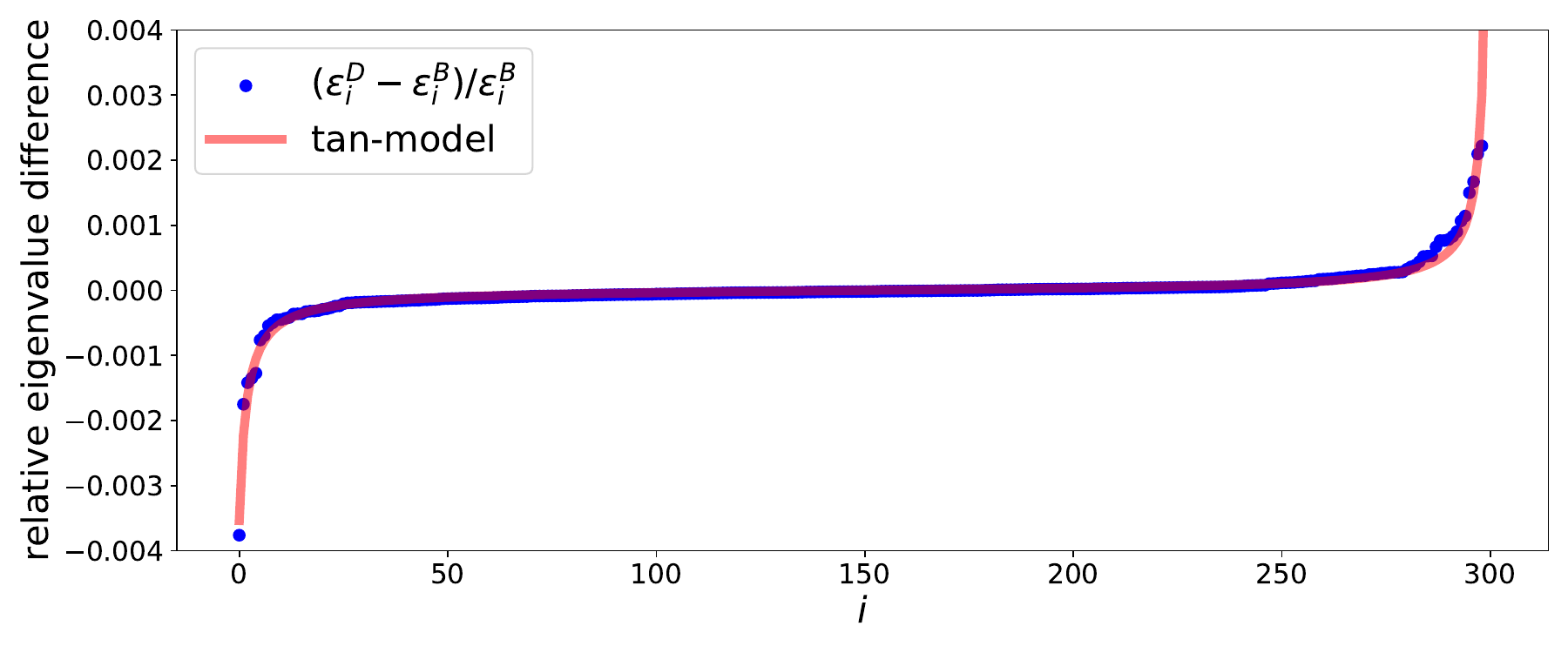}
\caption{VNet approximating the two-body interaction of the H$_6$ molecule. Top-left: Training and testing MAE of VNet across various geometries. Top-right: $(Z^B)^T W^D Z^B$ representation. Bottom: Empirical \texttt{tan}-model. 
}
\label{fig:h6twobody}
\end{figure*}

\begin{figure*}
\includegraphics[width=0.45\textwidth
]{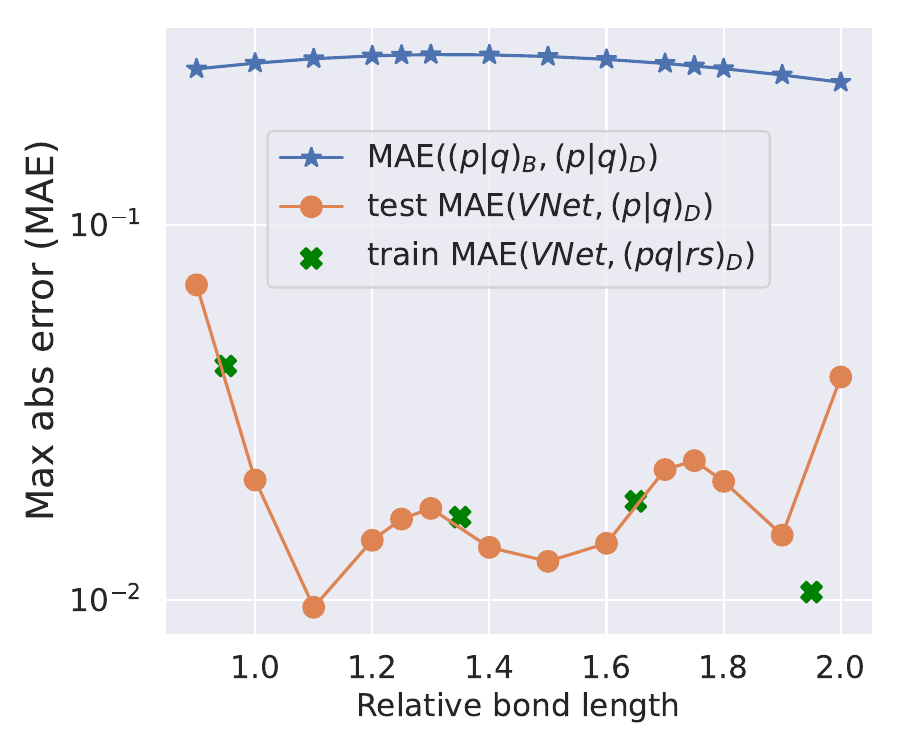}
\includegraphics[width=0.49\textwidth
]{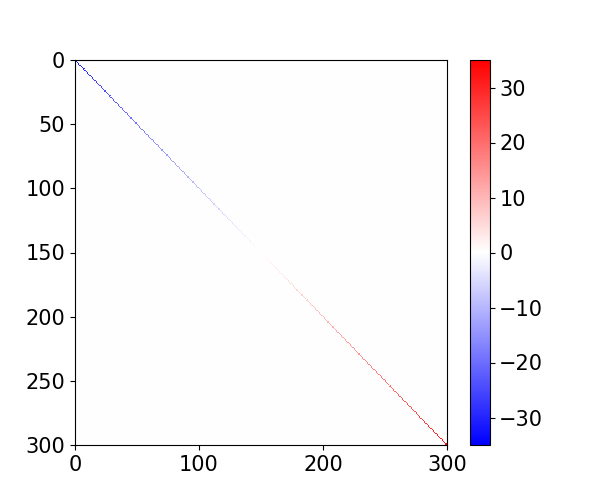}
\includegraphics[width=0.8\textwidth
]{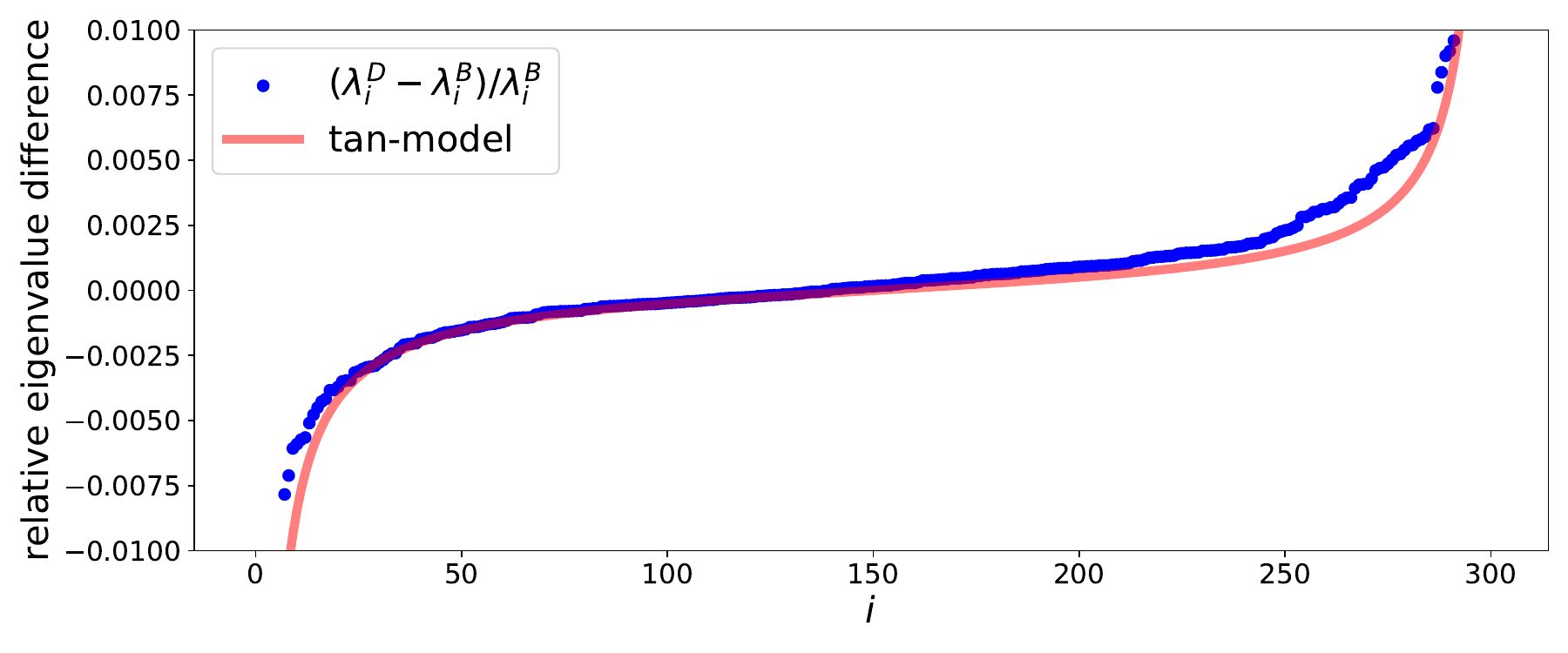}
\caption{VNet approximating the one-body interaction of the HF molecule. Top-left: Training and testing MAE of VNet across various geometries. Top-right: $(X^B)^T M^D X^B$ representation. Bottom: Empirical \texttt{tan}-model. 
}
\label{fig:HFonebody}
\end{figure*}

\begin{figure*}
\includegraphics[width=0.45\textwidth
]{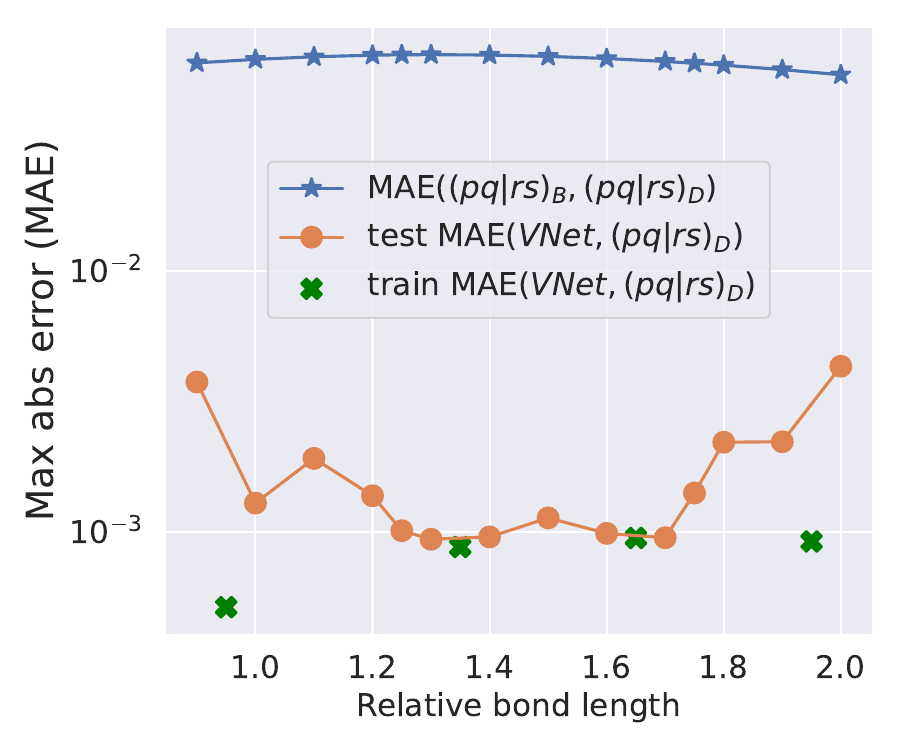}
\includegraphics[width=0.49\textwidth
]{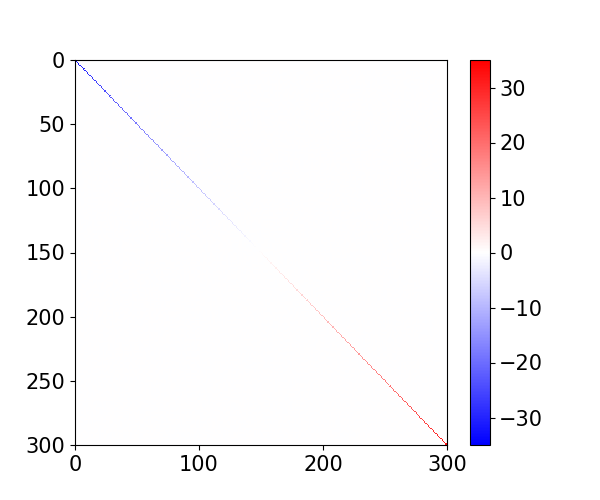}
\includegraphics[width=0.8\textwidth
]{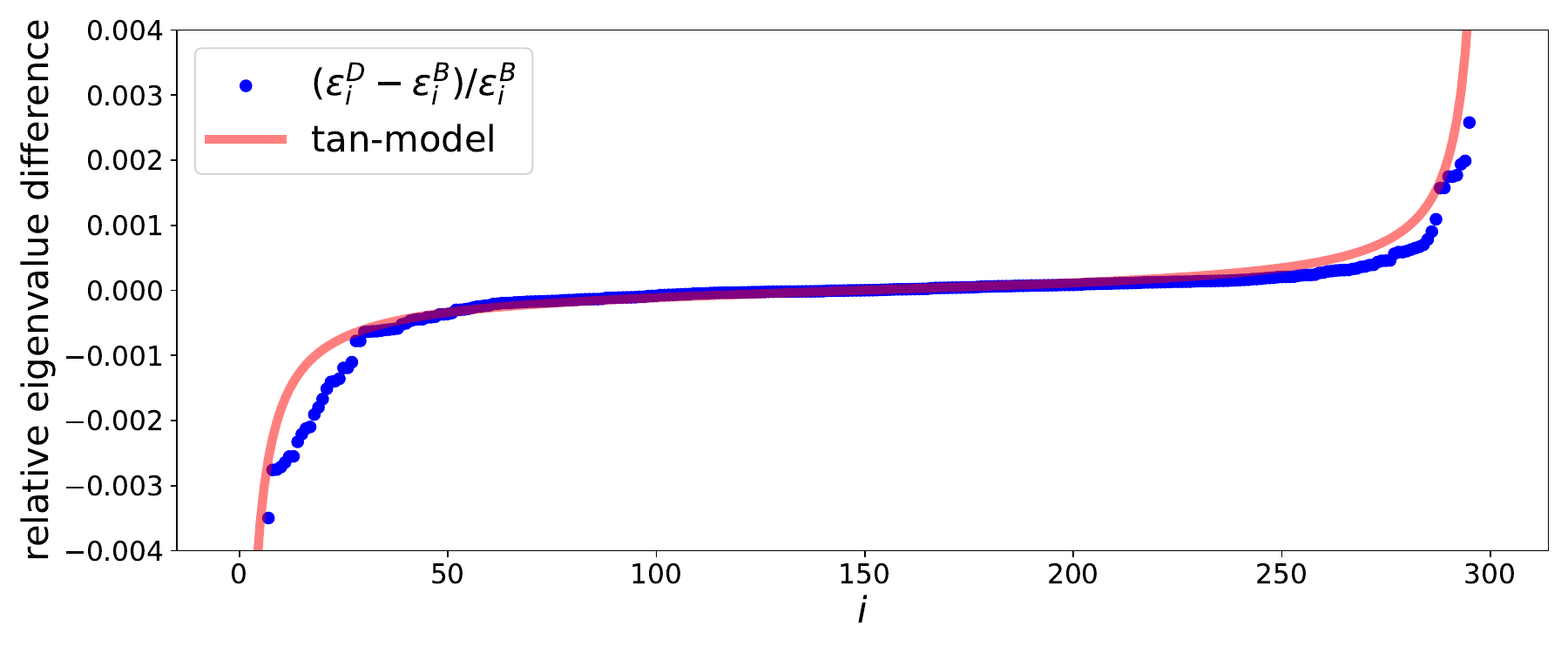}
\caption{VNet approximating the two-body interaction of the HF molecule. Top-left: Training and testing MAE of VNet across various geometries. Top-right: $(Z^B)^T W^D Z^B$ representation. Bottom: Empirical \texttt{tan}-model. 
}
\label{fig:HFtwobody}
\end{figure*}

\begin{figure*}
\includegraphics[width=0.45\textwidth
]{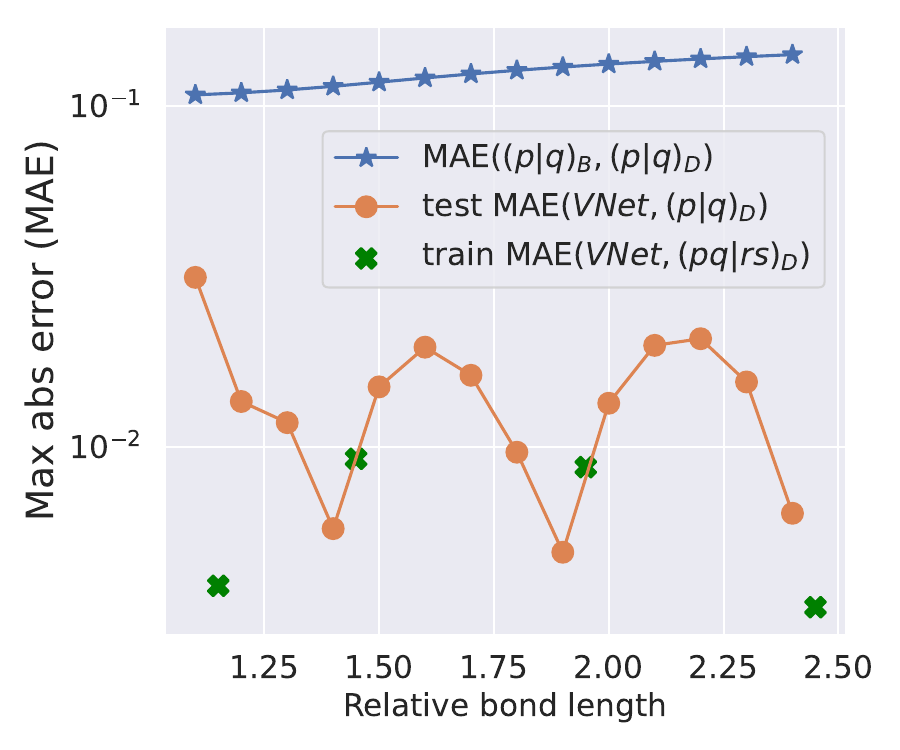}
\includegraphics[width=0.49\textwidth
]{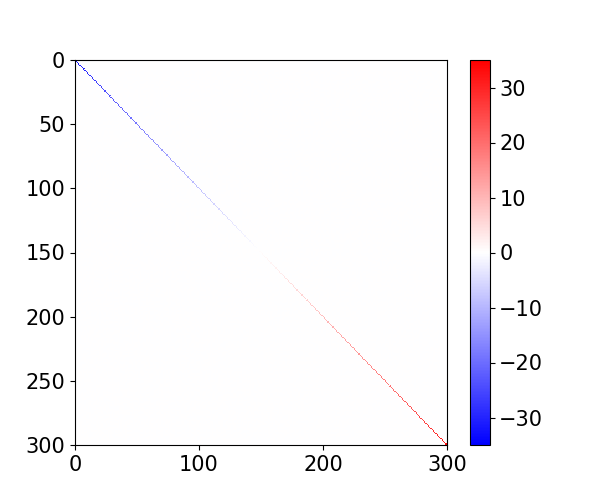}
\includegraphics[width=0.8\textwidth
]{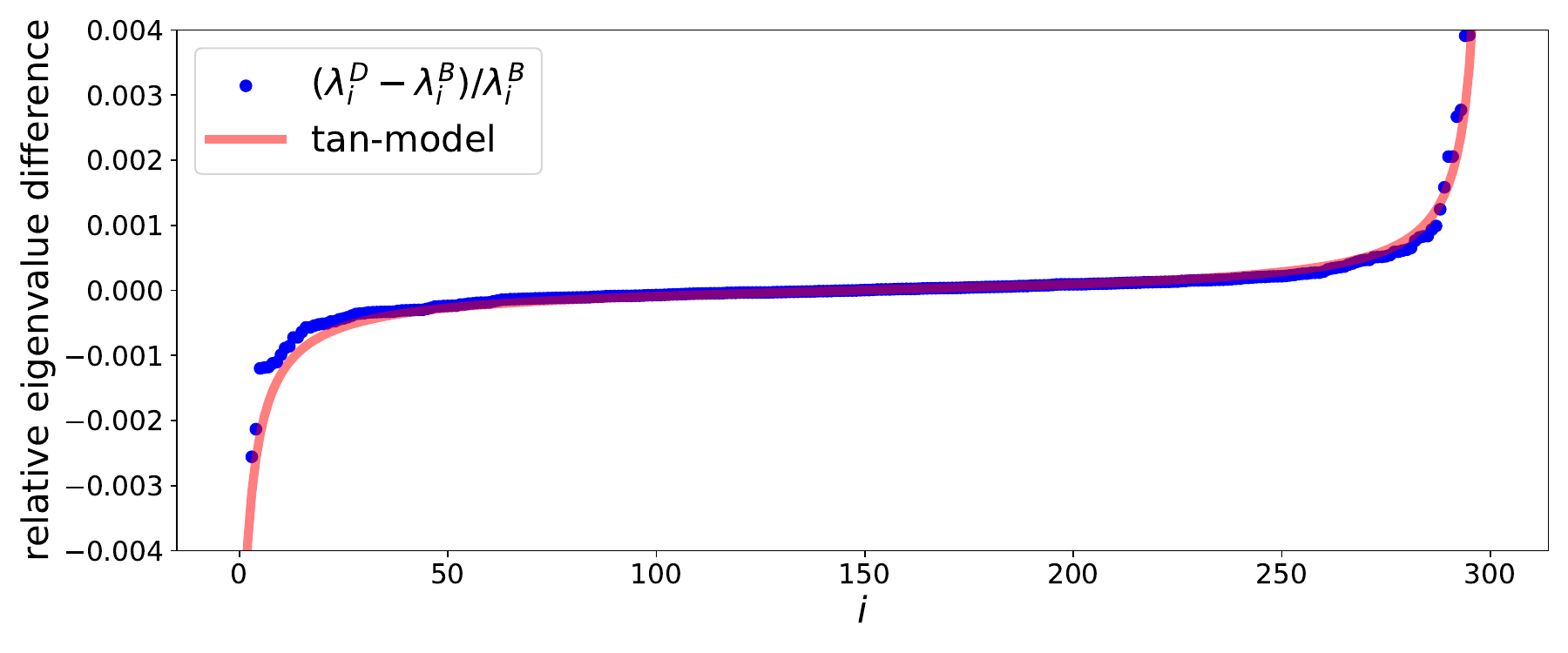}
\caption{VNet approximating the one-body interaction of the H$_2$O molecule. Top-left: Training and testing MAE of VNet across various geometries. Top-right: $(X^B)^T M^D X^B$ representation. Bottom: Empirical \texttt{tan}-model. 
}
\label{fig:H2Oonebody}
\end{figure*}

\begin{figure*}
\includegraphics[width=0.45\textwidth
]{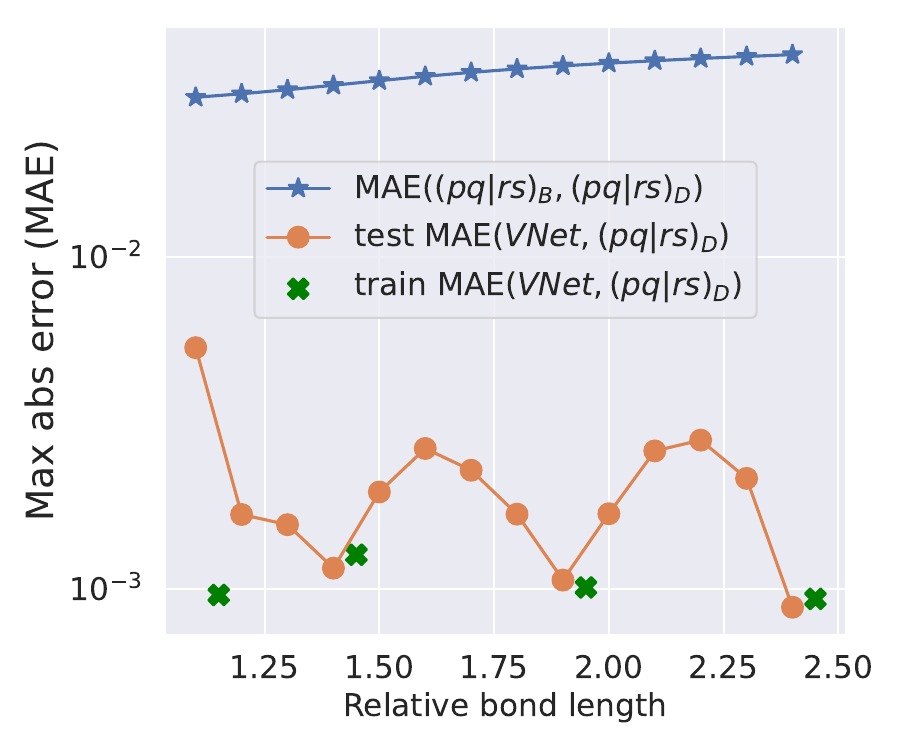}
\includegraphics[width=0.49\textwidth
]{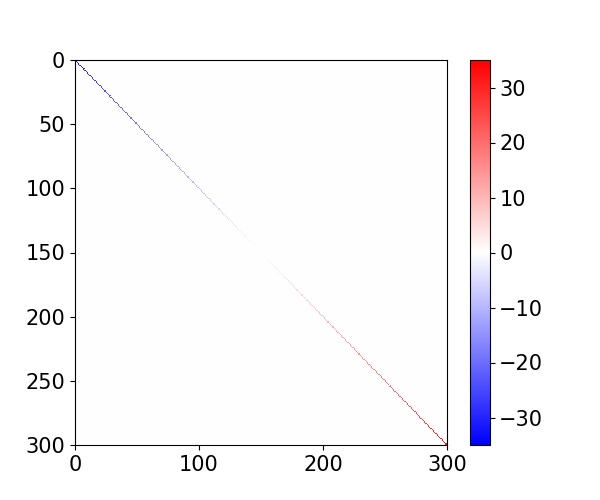}
\includegraphics[width=0.8\textwidth
]{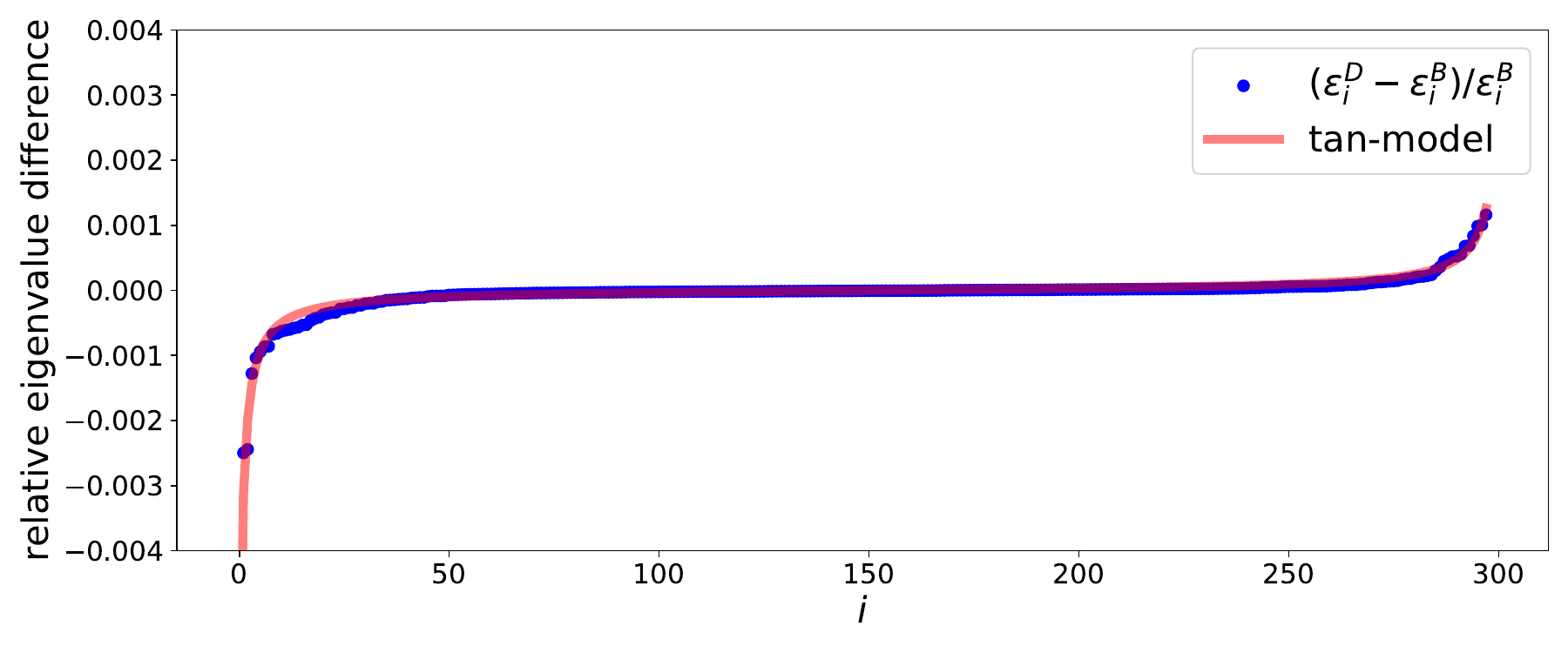}
\caption{VNet approximating the two-body interaction of the H$_2$O molecule. Top-left: Training and testing MAE of VNet across various geometries. Top-right: $(Z^B)^T W^D Z^B$ representation. Bottom: Empirical \texttt{tan}-model. 
}
\label{fig:H2Otwobody}
\end{figure*}

Figures~\ref{fig:h4onebody}, \ref{fig:h6onebody}, \ref{fig:HFonebody} and \ref{fig:H2Oonebody} present the one-body results for H$_4$, H$_6$, HF, and H$_2$O respectively, while Figures~\ref{fig:h4twobody}, \ref{fig:h6twobody}, \ref{fig:HFtwobody} and \ref{fig:H2Otwobody} display the two-body results. These results indicate that the MAE for the VNet ranges between $10^{-3}$ and $10^{-2}$ for one-body cases and is approximately $10^{-3}$ for two-body cases. The top-right subfigures show the matrix elements of $(Z^B)^T Z^D$ or $(X^B)^T X^D$ as a heatmap.  It is clear from these figures that the off-diagonal elements of $(Z^B)^T Z^D$ and $(X^B)^T X^D$ are tiny, which indicates that the eigenvectors of the interaction kernels learned from the bare tensors ($M^B$ or $W^B$) align closely with those of the effective interaction tensors ($M^D$ or $W^D$). The bottom subfigures present the relative eigenvalue differences and the empirically fitted \texttt{tan}-model. The coefficients of the empirical \texttt{tan}-model are provided in Table~\ref{tab:tancoeffs}. Interestingly, the learned coefficients are very similar (or of the same magnitude) across different systems, which may suggest an underlying relationship between the bare and effective interaction kernel tensors.


\begin{table}[htbp]
  \centering
  \caption{Coefficients in the \texttt{tan}-model for $\ell=300$. }
  \begin{adjustbox}{width=0.47\textwidth}
    \begin{tabular}{ccrrrrrrr}
    \toprule
          &       & \multicolumn{3}{c}{One-body} &       & \multicolumn{3}{c}{Two-body} \\
    \cmidrule{3-5}\cmidrule{7-9}
          &       & \multicolumn{1}{c}{$\gamma \ \ $}  & \multicolumn{1}{c}{$\delta \ \ $} &  \multicolumn{1}{c}{$i_e$}&       & \multicolumn{1}{c}{$\alpha$} & \multicolumn{1}{c}{$\beta$} &  \multicolumn{1}{c}{$i_c$}\\
\cmidrule{3-5}\cmidrule{7-9}    H$_4$    &     & $1.0\times 10^{-2}$    &   $1.0\times 10^{-4}$    &  150.2    &       &       $1.0\times 10^{-2}$       &    $0.6\times 10^{-4}$   &   149.8  \\
    H$_6$    &       &   $1.0\times 10^{-2}$    &   $3.1\times 10^{-4}$    &  147.8    &       &        $1.0\times 10^{-2}$    &    $0.6\times 10^{-4}$   &   149.1 \\
    HF    &       & $1.0\times 10^{-2}$    &   $8.9\times 10^{-4}$    &  150.4             &       &  $1.1\times 10^{-2}$       &    $2.0\times 10^{-4}$   &   149.4\\
    H$_2$O   &       &   $1.0\times 10^{-2}$   &   $1.6\times 10^{-4}$    &  148.4    &       &   $1.1\times 10^{-2}$    &   $0.6\times 10^{-4}$    &   150.3 \\
    \bottomrule
    \end{tabular}%
    \end{adjustbox}
  \label{tab:tancoeffs}%
\end{table}%

\section{Relationships between tangent-model and Suzuki effective interaction formalism}

%

In this section, we draw the connection between the empirically observed tangent dependencies  (Eqs.(\ref{tan2},\ref{tan21})) between tensors defining bare and dressed interactions and the previously developed theory that characterizes the relationship between original (bare) Hamiltonian and the effective Hamiltonian derived from a similarity transformation.

Methods for deriving an effective Hamiltonian through a similarity transformation have been widely used in chemistry and physics 
(see, for example, Refs.~\cite{van1929sigma,
jordahl1934effect,
bloch1958theorie,
des1960extension,
primas1961verallgemeinerte,
primas1963generalized,
lowdin1963studies,
schrieffer1966relation,
kirtman1968variational,
soliverez1969effective,
jorgensen1974projector,
jorgensen1975effective,
mukherjee1977ab,
lindgren1978coupled,
lindgren2012atomic,
shavitt1980quasidegenerate,
westhaus1981connections,
suzuki1982construction,
suzuki1983degenerate,
glazek1993renormalization,
bravyi2011schrieffer,
stroberg2019nonempirical,
durand1983direct,
durand1987effective,
jezmonk,
pal1988molecular,
jeziorski1989valence,
kaldor1991fock,
bernholdt1999critical,
meissner1998fock,
evangelista2007coupling,
mrcclyakh}).
Several formulations, originating in the canonical formulation of  Van Vleck transformation \cite{van1929sigma}, have been independently developed in the areas of quantum chemistry and nuclear structure theory to describe quasi-degenerate states. 
For example, Shavitt in Ref.~\cite{shavitt1980quasidegenerate} and Suzuki in Ref.~\cite{suzuki1982construction} came across similar multi-reference formalisms based on the unitary transformations defined by the exponents of the anti-Hermitian cluster operators. In both approaches, the properties of anti-Hermitian cluster operators were used to characterize the relationships between effective Hamiltonians and bare Hamiltonians in the corresponding active (or model) spaces (see also Ref.~\cite{hoffmann1996canonical}). 
Our analysis is based on the theoretical results discussed in Refs.~\cite{shavitt1980quasidegenerate,suzuki1982construction}. 

Following the original analysis \cite{shavitt1980quasidegenerate,suzuki1982construction}, let us consider a unitary transformation of the original Hamiltonian $H$ 
\begin{equation}
    \bar{H} = e^{-G} H e^G \;
    \label{dis1}
\end{equation}
where anti-Hermitian cluster operator $G$ ($G^{\dagger} = -G$)
is subject to the condition 
\begin{equation}
Q (e^{-G} H e^G) R = R (e^{-G} H e^G) Q = 0 \;,
\label{dis2}
\end{equation}
where the projection operators $Q$ and $R$ are defined as
\begin{eqnarray}
    R &=& P+Q_{\rm int}({\mathfrak{h}}) \;, \label{dis3} \\
    Q &=& Q_{\rm ext}({\mathfrak{h}}) \;, \label{diis3b}\\
    1 &=& R + Q  \;. \label{dis4}
\end{eqnarray}
The so-called minimal effect requirement 
\begin{equation}
    RGR = QGQ = 0 \; \label{dis5}  
\end{equation}
is imposed. The above conditions leads to the effective Hamiltonian $H^{\rm eff}_S$ defined as
\begin{equation}
H^{\rm eff}_S = R e^{-G} H e^{G} R \;. 
\label{dis6}
\end{equation}
The many-body analysis of Ref.~\cite{suzuki1982construction}
provides an estimate of the difference between the effective $H^{\rm eff}_S$ Hamiltonian and a bare Hamiltonian $H$ in the active space $R$, denoted by
 \begin{equation}
W = H^{\rm eff}_S - R H R.
 \label{dis7}
 \end{equation}
It has been shown that the $W$ operator takes the form
\begin{equation}
W = -R\lbrace Z_{\rm tanh} [G/2,H_{\rm OD}] \rbrace R,
\label{dis8}
\end{equation}
where $H_{\rm OD}$ is defined as
\begin{equation}
H_{\rm OD} = RHQ + QHR \;,
\label{dis9}
\end{equation}    
and $Z_{\rm tanh}$ is a commutator function. The coefficients in the Taylor expansion of $Z_{\rm tanh}$ correspond to those in the Taylor expansion
of the $\tanh$ function, i.e.,
\begin{widetext}
\begin{equation}
Z_{\rm tanh} [G/2,H_{\rm OD}] = \sum_{n=1}^{\infty}
\lbrace (-1)^{n-1} 2^{2n} (2^{2n}-1) B_n/(2n)!  \rbrace
[(G/2)^{\lbrace 2n-1 \rbrace},H_{\rm OD}] \;,
\label{dis10}
\end{equation}
\end{widetext}
where notation $[A^{\lbrace i \rbrace},B]$ is a $i$-tuple commutator expression
\begin{equation}
    [A^{\lbrace i \rbrace},B] = [A,[A,[...[A,B]...]]]\;.
    \label{dis11}
\end{equation}

To evaluate the explicit algebraic form of the expansion (\ref{dis10}), let us introduce a basis set $\lbrace |\chi_k\rangle \rbrace_{k=1}^M$ spanned by eigenvectors of $G/2$ matrix, 
\begin{equation}
 (G/2) |\chi_k \rangle = 
 \gamma_k |\chi_k\rangle \;\;, (k=1,
 \ldots,M) \;,
 \label{dis12}
\end{equation}
where $\gamma_k$ is purely imaginary, i.e, $\gamma_k= i g_k$
where $g_k$ is a real number. The matrix representation of the Hermitian $H_{\rm OD}$ in the $\lbrace |\chi_k\rangle \rbrace_{k=1}^M$ basis is defined by matrix elements $h_{kl}$
\begin{equation}
H_{\rm OD} \rightarrow [h_{kl}^{\rm OD}] \;.
\label{dis13}
\end{equation}
Now, let us evaluate matrix elements of $C^{\lbrace i \rbrace }=[A^{\lbrace i \rbrace},B] = [A,[A,[...[A,B]...]]]$ in eigenbasis of operator $A$ with the corresponding eigenvalues $\alpha_k=i a_k$, where $a_k$ is real. In this basis, the matrix elements correpsonding to operators $A$ and $B$ are defined as $[\delta_{kl}\alpha_{l}]$ and $b_{kl}$, respectively.
Using the recursion 
\begin{equation}
C^{\lbrace i \rbrace } = [A,C^{\lbrace i-1 \rbrace }]
\label{dis14}
\end{equation}
and the fact that 
\begin{equation}
C^{\lbrace 1 \rbrace } \rightarrow [b_{kl}(a_k-a_l)] \;,
\label{dis15}
\end{equation}
one can show that matrix elements of $C^{\lbrace i \rbrace }$ are defined as 
\begin{equation}
C^{\lbrace i \rbrace } \rightarrow [b_{kl}(a_k-a_l)^i] \;.
\label{dis16}
\end{equation}
Applying the above identity to Eq.(\ref{dis10}) leads to the following expression
\begin{equation}
Z_{\rm tanh} [G/2,H_{\rm OD}] \rightarrow [h_{kl}^{\rm OD} \tanh(\gamma_k-\gamma_l)] \;.
\label{dis17}
\end{equation}
Using the following identity 
\begin{equation}
\tanh(ia) = -i \tan(a)
\label{dis18}
\end{equation}
for any $a \in \mathbb{R}$,
we can rewrite (\ref{dis17}) as
\begin{equation}
Z_{\rm tanh} [G/2,H_{\rm OD}] \rightarrow 
i [h_{kl}^{\rm OD} \tan(g_k-g_l)] \;,
\label{dis19}
\end{equation}
where $g_k = -i\gamma_k \in \mathbb{R}$. 
The above expression indicates that $Z_{\rm tanh} [G/2,H_{\rm OD}]$, which provides an estimate of the difference between the bare and effective Hamiltonian in an active space, 
are related to the matrix elements of the bare Hamiltonian by a tangent function. This relationship bears a remarkable resemblance to the relationship between the eigenvalues of the bare and effective interaction kernels revealed by the VNet representation of these interactions.

The eigenvalues of the anti-symmetric matrix $[h_{kl}^{\rm OD} \tan(g_k-g_l)]$ are given by purely imaginary numbers $\zeta_j= i z_j$'s, where $z_k$ is a real number. Therefore, eigenvalues of $Z_{\rm tanh} [G/2,H_{\rm OD}]$ are given by real numbers $z_j$'s, which are nontrivial functions of $\tan (g_k-g_l)$'s. 

Although it is challenging to show how the tangent relationship between the bare and effective interaction kernel \eqref{tan2} and \eqref{tan21} directly translates into the tangent relationship between the bare and effective Hamiltonians exhibited by \eqref{dis19}, it is clear that the origin of the tangent function in \eqref{dis19} lies in the presence of the tangent function in \eqref{tan2} and \eqref{tan21}, i.e., when the bare and effective interactions are used in the second quantized representation of the corresponding Hamiltonians, the tangent function present in \eqref{tan2} and \eqref{tan21} will appear in some form in the difference of these two Hamiltonians.

\begin{figure*}
\includegraphics[width=0.8\textwidth
]{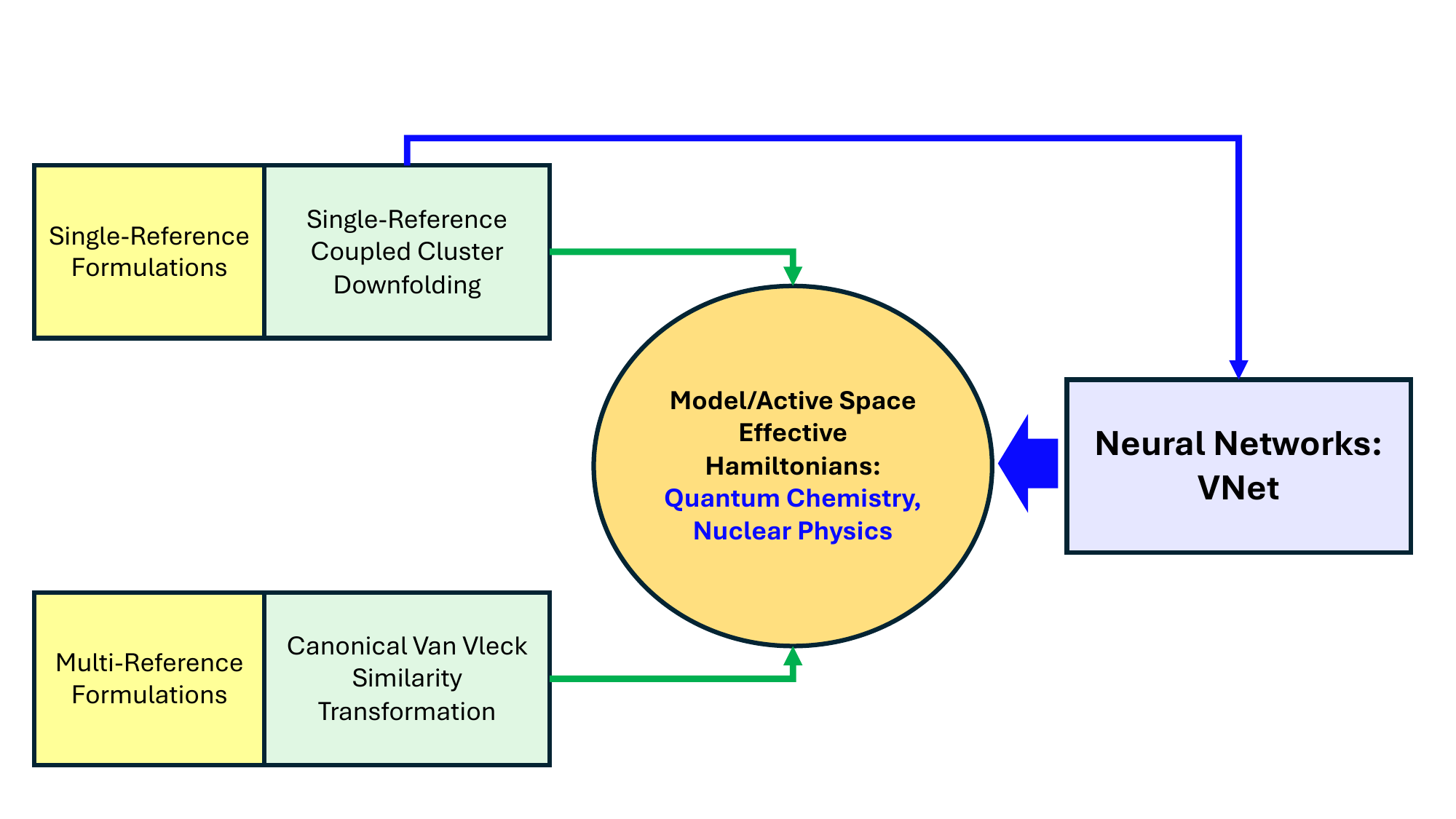}
\caption{Machine Learning extraction of properties of the genuine multi-reference CC approaches using Neural Networks and single-reference Hermitian CC downfolding methods.
}
\label{fig:vanv}
\end{figure*}

The preceding analysis shows that the VNet representation of the effective one and two-body interactions used in SR-CC-based downfolding methodologies (see Fig.~\ref{fig:vanv}) not only provides an efficient way to evaluate these interactions to enable practical use of SR-CC formulation to circumvent certain limitations associated with multi-reference CC theories, but also reveals the nature of effective interaction that yields additional insight into the previously developed theory that characterizes the difference between the effective Hamiltonian and the bare Hamiltonian. 



\begin{table*}[htbp]
  \centering
  \caption{VNet training configurations for one-body interaction tensor.}
    \begin{tabular}{cccccccccc}
    \toprule
          &       & \multicolumn{3}{c}{Bare tensor training} &       &       & \multicolumn{3}{c}{Effective tensor training} \\
    \cmidrule{3-5}\cmidrule{8-10}
          &       & \#Epoch & \multicolumn{1}{l}{Batch size} & Learning rate &       &       & \#Epoch & \multicolumn{1}{l}{Batch size} & Learning rate \\
\cmidrule{3-5}\cmidrule{8-10}    H$_4$    &       & 8000  & 128   & 0.001 &       &       & 500   & 2048  & 0.001 \\
    H$_6$    &   & 8000    &    128   &    0.001        &       &       &  500     &    2048   &  0.0001 \\
    HF    &       &    8000   & 1024      &   0.002    &       &       &   500    &  512     & 0.02 \\
    H$_2$O   &       &   2000    & 256      &   0.0002    &       &       &    500   &  1024     & 0.001 \\
    \bottomrule
    \end{tabular}%
  \label{tab:Onebodytraining}%
\end{table*}%

\begin{table*}[htbp]
  \centering
  \caption{VNet training configurations for two-body interaction tensor.}
    \begin{tabular}{cccccccccc}
    \toprule
          &       & \multicolumn{3}{c}{Bare tensor training} &       &       & \multicolumn{3}{c}{Effective tensor training} \\
    \cmidrule{3-5}\cmidrule{8-10}
          &       & \#Epoch & \multicolumn{1}{l}{Batch size} & Learning rate &       &       & \#Epoch & \multicolumn{1}{l}{Batch size} & Learning rate \\
\cmidrule{3-5}\cmidrule{8-10}    H$_4$    &       & 2000  & 256   & 0.001 &       &       & 500   & 128  & 0.0002 \\
    H$_6$    &            &    2000   &   1024    &   0.001  &   &   & 500    &    128         &  0.0002 \\
    HF    &       &    2000   &    1024   &   0.001    &       &       &   500    &   512    & 0.0002 \\
   H$_2$O   &       &  5000     &  1024     &   0.001    &       &       &   500    &  4096     &  0.0002\\
    \bottomrule
    \end{tabular}%
  \label{tab:Twobodytraining}%
\end{table*}%

\section{Conclusion}

The VNet methodology exemplifies the application of neural networks for identifying common features of many-body formulations that arise from distinct theoretical frameworks stemming from single-reference and multi-reference coupled-cluster theories. These shared characteristics manifested by the \texttt{tan}-model, intrinsically linked to the structure of effective Hamiltonians and explored herein through NN techniques, may offer alternative approaches for investigating strongly correlated systems 
instead of coping with challenges associated with the inclusion of high-rank excitations that impose substantial computational limitations and issues related to convergence or stability of the underlying expansions for the CC wave function parametrization. 

Moreover, because the \texttt{tan}-model contains only a few parameters that can be used to map a bare interaction to an effective interaction, it presents a potential pathway towards constructing reduced-scaling models applicable to a broad spectrum of problems, where effective Hamiltonians can be tailored/trained to specific scenarios characterized by suitably defined active spaces, such as those involved in single, double, and triple bond-breaking processes. The current analysis also suggests that there is flexibility in defining computational kernels that characterize bare and dressed interactions, which remain subjects of ongoing research. 

The numerical analysis of the  $M^B$/$M^D$ and $W^B$/$W^D$ kernels can be summarized as follows:
\begin{itemize}
\item For all cases considered here, all eigenvectors of $M^B$ and  $W^B$ nearly diagonalize the $M^D$ and  $W^D$ matrices, respectively.
\item The eigenvectors of $M^B$ and $W^B$ can also be used to characterize the difference between bare and downfolded kernels, i.e., $(M^D-M^B)$ and $(W^D-W^B)$.
\item Both $M^D$ and $W^D$ kernels disclose similar tangent-type behavior defined by Eqs.
(\ref{tan2}) and (\ref{tan21}).
\end{itemize}
These properties allows us to examine the relationship between the bare and effective interactions by comparing the eigenvalues of the interaction kernel matrices produced from the VNet representation. We found that the emerged relationship between these eigenvalues exhibited by the \texttt{tan}-model has a close connection with the Suzuki formalism and aligns with the analysis of the most dominant effects defining correlation energy.

\section{acknowledgments}
This material is based upon work supported by the ``Transferring exascale computational chemistry to cloud computing environment and emerging hardware technologies (TEC$^4$)''  project, which is funded by the U.S. Department of Energy, Office of Science, Office of Basic Energy Sciences, the Division of Chemical Sciences, Geosciences, and Biosciences (under FWP 82037). N.P.B. also acknowledges support from the Laboratory Directed Research and Development (LDRD) Program at Pacific Northwest National Laboratory. 
This work is also supported by the U.S. Department of Energy, Office of Science, Office of Advanced Scientific Computing Research and Office of Basic Energy Science, Scientific Discovery through Advanced Computing (SciDAC) program under Contract No. DE-AC02-05CH11231. (S.L. and C.Y.) This work used the computational resources of the National Energy Research Scientific Computing (NERSC) center under NERSC Award ASCR-ERCAP m1027 for 2024, which is supported by the Office of Science of the U.S. Department of Energy under Contract No. DE-AC02-05CH11231.




\bibliography{apssamp}

\providecommand{\noopsort}[1]{}\providecommand{\singleletter}[1]{#1}%
\begin{thebibliography}{76}%
\makeatletter
\providecommand \@ifxundefined [1]{%
 \@ifx{#1\undefined}
}%
\providecommand \@ifnum [1]{%
 \ifnum #1\expandafter \@firstoftwo
 \else \expandafter \@secondoftwo
 \fi
}%
\providecommand \@ifx [1]{%
 \ifx #1\expandafter \@firstoftwo
 \else \expandafter \@secondoftwo
 \fi
}%
\providecommand \natexlab [1]{#1}%
\providecommand \enquote  [1]{``#1''}%
\providecommand \bibnamefont  [1]{#1}%
\providecommand \bibfnamefont [1]{#1}%
\providecommand \citenamefont [1]{#1}%
\providecommand \href@noop [0]{\@secondoftwo}%
\providecommand \href [0]{\begingroup \@sanitize@url \@href}%
\providecommand \@href[1]{\@@startlink{#1}\@@href}%
\providecommand \@@href[1]{\endgroup#1\@@endlink}%
\providecommand \@sanitize@url [0]{\catcode `\\12\catcode `\$12\catcode `\&12\catcode `\#12\catcode `\^12\catcode `\_12\catcode `\%12\relax}%
\providecommand \@@startlink[1]{}%
\providecommand \@@endlink[0]{}%
\providecommand \url  [0]{\begingroup\@sanitize@url \@url }%
\providecommand \@url [1]{\endgroup\@href {#1}{\urlprefix }}%
\providecommand \urlprefix  [0]{URL }%
\providecommand \Eprint [0]{\href }%
\providecommand \doibase [0]{https://doi.org/}%
\providecommand \selectlanguage [0]{\@gobble}%
\providecommand \bibinfo  [0]{\@secondoftwo}%
\providecommand \bibfield  [0]{\@secondoftwo}%
\providecommand \translation [1]{[#1]}%
\providecommand \BibitemOpen [0]{}%
\providecommand \bibitemStop [0]{}%
\providecommand \bibitemNoStop [0]{.\EOS\space}%
\providecommand \EOS [0]{\spacefactor3000\relax}%
\providecommand \BibitemShut  [1]{\csname bibitem#1\endcsname}%
\let\auto@bib@innerbib\@empty
\bibitem [{\citenamefont {Shinaoka}\ \emph {et~al.}(2015)\citenamefont {Shinaoka}, \citenamefont {Troyer},\ and\ \citenamefont {Werner}}]{shinaoka2015accuracy}%
  \BibitemOpen
  \bibfield  {author} {\bibinfo {author} {\bibfnamefont {H.}~\bibnamefont {Shinaoka}}, \bibinfo {author} {\bibfnamefont {M.}~\bibnamefont {Troyer}},\ and\ \bibinfo {author} {\bibfnamefont {P.}~\bibnamefont {Werner}},\ }\bibfield  {title} {\bibinfo {title} {Accuracy of downfolding based on the constrained random-phase approximation},\ }\href@noop {} {\bibfield  {journal} {\bibinfo  {journal} {Physical Review B}\ }\textbf {\bibinfo {volume} {91}},\ \bibinfo {pages} {245156} (\bibinfo {year} {2015})}\BibitemShut {NoStop}%
\bibitem [{\citenamefont {Aryasetiawan}\ and\ \citenamefont {Nilsson}(2022)}]{aryasetiawan2022downfolding}%
  \BibitemOpen
  \bibfield  {author} {\bibinfo {author} {\bibfnamefont {F.}~\bibnamefont {Aryasetiawan}}\ and\ \bibinfo {author} {\bibfnamefont {F.}~\bibnamefont {Nilsson}},\ }\href@noop {} {\emph {\bibinfo {title} {Downfolding methods in many-electron theory}}}\ (\bibinfo  {publisher} {AIP Publishing Melville, NY},\ \bibinfo {year} {2022})\BibitemShut {NoStop}%
\bibitem [{\citenamefont {Zheng}\ \emph {et~al.}(2018)\citenamefont {Zheng}, \citenamefont {Changlani}, \citenamefont {Williams}, \citenamefont {Busemeyer},\ and\ \citenamefont {Wagner}}]{zheng2018real}%
  \BibitemOpen
  \bibfield  {author} {\bibinfo {author} {\bibfnamefont {H.}~\bibnamefont {Zheng}}, \bibinfo {author} {\bibfnamefont {H.~J.}\ \bibnamefont {Changlani}}, \bibinfo {author} {\bibfnamefont {K.~T.}\ \bibnamefont {Williams}}, \bibinfo {author} {\bibfnamefont {B.}~\bibnamefont {Busemeyer}},\ and\ \bibinfo {author} {\bibfnamefont {L.~K.}\ \bibnamefont {Wagner}},\ }\bibfield  {title} {\bibinfo {title} {From real materials to model hamiltonians with density matrix downfolding},\ }\href@noop {} {\bibfield  {journal} {\bibinfo  {journal} {Frontiers in Physics}\ }\textbf {\bibinfo {volume} {6}},\ \bibinfo {pages} {43} (\bibinfo {year} {2018})}\BibitemShut {NoStop}%
\bibitem [{\citenamefont {Chang}\ \emph {et~al.}(2024)\citenamefont {Chang}, \citenamefont {van Loon}, \citenamefont {Eskridge}, \citenamefont {Busemeyer}, \citenamefont {Morales}, \citenamefont {Dreyer}, \citenamefont {Millis}, \citenamefont {Zhang}, \citenamefont {Wehling}, \citenamefont {Wagner} \emph {et~al.}}]{chang2024downfolding}%
  \BibitemOpen
  \bibfield  {author} {\bibinfo {author} {\bibfnamefont {Y.}~\bibnamefont {Chang}}, \bibinfo {author} {\bibfnamefont {E.~G.}\ \bibnamefont {van Loon}}, \bibinfo {author} {\bibfnamefont {B.}~\bibnamefont {Eskridge}}, \bibinfo {author} {\bibfnamefont {B.}~\bibnamefont {Busemeyer}}, \bibinfo {author} {\bibfnamefont {M.~A.}\ \bibnamefont {Morales}}, \bibinfo {author} {\bibfnamefont {C.~E.}\ \bibnamefont {Dreyer}}, \bibinfo {author} {\bibfnamefont {A.~J.}\ \bibnamefont {Millis}}, \bibinfo {author} {\bibfnamefont {S.}~\bibnamefont {Zhang}}, \bibinfo {author} {\bibfnamefont {T.~O.}\ \bibnamefont {Wehling}}, \bibinfo {author} {\bibfnamefont {L.~K.}\ \bibnamefont {Wagner}}, \emph {et~al.},\ }\bibfield  {title} {\bibinfo {title} {Downfolding from ab initio to interacting model hamiltonians: comprehensive analysis and benchmarking of the dft+ crpa approach},\ }\href@noop {} {\bibfield  {journal} {\bibinfo  {journal} {npj Computational Materials}\ }\textbf {\bibinfo {volume} {10}},\ \bibinfo {pages} {129} (\bibinfo
  {year} {2024})}\BibitemShut {NoStop}%
\bibitem [{\citenamefont {Romanova}\ \emph {et~al.}(2023)\citenamefont {Romanova}, \citenamefont {Weng}, \citenamefont {Apelian},\ and\ \citenamefont {Vlcek}}]{romanova2023dynamical}%
  \BibitemOpen
  \bibfield  {author} {\bibinfo {author} {\bibfnamefont {M.}~\bibnamefont {Romanova}}, \bibinfo {author} {\bibfnamefont {G.}~\bibnamefont {Weng}}, \bibinfo {author} {\bibfnamefont {A.}~\bibnamefont {Apelian}},\ and\ \bibinfo {author} {\bibfnamefont {V.}~\bibnamefont {Vlcek}},\ }\bibfield  {title} {\bibinfo {title} {Dynamical downfolding for localized quantum states. npj comput},\ }\href@noop {} {\bibfield  {journal} {\bibinfo  {journal} {Mater}\ }\textbf {\bibinfo {volume} {9}},\ \bibinfo {pages} {1} (\bibinfo {year} {2023})}\BibitemShut {NoStop}%
\bibitem [{\citenamefont {Georges}\ \emph {et~al.}(1996)\citenamefont {Georges}, \citenamefont {Kotliar}, \citenamefont {Krauth},\ and\ \citenamefont {Rozenberg}}]{georges1996dynamical}%
  \BibitemOpen
  \bibfield  {author} {\bibinfo {author} {\bibfnamefont {A.}~\bibnamefont {Georges}}, \bibinfo {author} {\bibfnamefont {G.}~\bibnamefont {Kotliar}}, \bibinfo {author} {\bibfnamefont {W.}~\bibnamefont {Krauth}},\ and\ \bibinfo {author} {\bibfnamefont {M.~J.}\ \bibnamefont {Rozenberg}},\ }\bibfield  {title} {\bibinfo {title} {Dynamical mean-field theory of strongly correlated fermion systems and the limit of infinite dimensions},\ }\href@noop {} {\bibfield  {journal} {\bibinfo  {journal} {Reviews of Modern Physics}\ }\textbf {\bibinfo {volume} {68}},\ \bibinfo {pages} {13} (\bibinfo {year} {1996})}\BibitemShut {NoStop}%
\bibitem [{\citenamefont {Kotliar}\ \emph {et~al.}(2006)\citenamefont {Kotliar}, \citenamefont {Savrasov}, \citenamefont {Haule}, \citenamefont {Oudovenko}, \citenamefont {Parcollet},\ and\ \citenamefont {Marianetti}}]{kotliar2006electronic}%
  \BibitemOpen
  \bibfield  {author} {\bibinfo {author} {\bibfnamefont {G.}~\bibnamefont {Kotliar}}, \bibinfo {author} {\bibfnamefont {S.~Y.}\ \bibnamefont {Savrasov}}, \bibinfo {author} {\bibfnamefont {K.}~\bibnamefont {Haule}}, \bibinfo {author} {\bibfnamefont {V.~S.}\ \bibnamefont {Oudovenko}}, \bibinfo {author} {\bibfnamefont {O.}~\bibnamefont {Parcollet}},\ and\ \bibinfo {author} {\bibfnamefont {C.}~\bibnamefont {Marianetti}},\ }\bibfield  {title} {\bibinfo {title} {Electronic structure calculations with dynamical mean-field theory},\ }\href@noop {} {\bibfield  {journal} {\bibinfo  {journal} {Reviews of Modern Physics}\ }\textbf {\bibinfo {volume} {78}},\ \bibinfo {pages} {865} (\bibinfo {year} {2006})}\BibitemShut {NoStop}%
\bibitem [{\citenamefont {Gull}\ \emph {et~al.}(2011)\citenamefont {Gull}, \citenamefont {Millis}, \citenamefont {Lichtenstein}, \citenamefont {Rubtsov}, \citenamefont {Troyer},\ and\ \citenamefont {Werner}}]{gull2011continuous}%
  \BibitemOpen
  \bibfield  {author} {\bibinfo {author} {\bibfnamefont {E.}~\bibnamefont {Gull}}, \bibinfo {author} {\bibfnamefont {A.~J.}\ \bibnamefont {Millis}}, \bibinfo {author} {\bibfnamefont {A.~I.}\ \bibnamefont {Lichtenstein}}, \bibinfo {author} {\bibfnamefont {A.~N.}\ \bibnamefont {Rubtsov}}, \bibinfo {author} {\bibfnamefont {M.}~\bibnamefont {Troyer}},\ and\ \bibinfo {author} {\bibfnamefont {P.}~\bibnamefont {Werner}},\ }\bibfield  {title} {\bibinfo {title} {Continuous-time monte carlo methods for quantum impurity models},\ }\href@noop {} {\bibfield  {journal} {\bibinfo  {journal} {Reviews of Modern Physics}\ }\textbf {\bibinfo {volume} {83}},\ \bibinfo {pages} {349} (\bibinfo {year} {2011})}\BibitemShut {NoStop}%
\bibitem [{\citenamefont {Zgid}\ and\ \citenamefont {Chan}(2011)}]{zgid2011dynamical}%
  \BibitemOpen
  \bibfield  {author} {\bibinfo {author} {\bibfnamefont {D.}~\bibnamefont {Zgid}}\ and\ \bibinfo {author} {\bibfnamefont {G.~K.}\ \bibnamefont {Chan}},\ }\bibfield  {title} {\bibinfo {title} {Dynamical mean-field theory from a quantum chemical perspective},\ }\href@noop {} {\bibfield  {journal} {\bibinfo  {journal} {The Journal of chemical physics}\ }\textbf {\bibinfo {volume} {134}} (\bibinfo {year} {2011})}\BibitemShut {NoStop}%
\bibitem [{\citenamefont {Lan}\ \emph {et~al.}(2015)\citenamefont {Lan}, \citenamefont {Kananenka},\ and\ \citenamefont {Zgid}}]{lan2015communication}%
  \BibitemOpen
  \bibfield  {author} {\bibinfo {author} {\bibfnamefont {T.~N.}\ \bibnamefont {Lan}}, \bibinfo {author} {\bibfnamefont {A.~A.}\ \bibnamefont {Kananenka}},\ and\ \bibinfo {author} {\bibfnamefont {D.}~\bibnamefont {Zgid}},\ }\bibfield  {title} {\bibinfo {title} {Communication: Towards ab initio self-energy embedding theory in quantum chemistry},\ }\href@noop {} {\bibfield  {journal} {\bibinfo  {journal} {The Journal of chemical physics}\ }\textbf {\bibinfo {volume} {143}} (\bibinfo {year} {2015})}\BibitemShut {NoStop}%
\bibitem [{\citenamefont {Knizia}\ and\ \citenamefont {Chan}(2012)}]{knizia2012density}%
  \BibitemOpen
  \bibfield  {author} {\bibinfo {author} {\bibfnamefont {G.}~\bibnamefont {Knizia}}\ and\ \bibinfo {author} {\bibfnamefont {G.~K.-L.}\ \bibnamefont {Chan}},\ }\bibfield  {title} {\bibinfo {title} {Density matrix embedding: A simple alternative to dynamical mean-field theory},\ }\href@noop {} {\bibfield  {journal} {\bibinfo  {journal} {Physical review letters}\ }\textbf {\bibinfo {volume} {109}},\ \bibinfo {pages} {186404} (\bibinfo {year} {2012})}\BibitemShut {NoStop}%
\bibitem [{\citenamefont {Knizia}\ and\ \citenamefont {Chan}(2013)}]{knizia2013density}%
  \BibitemOpen
  \bibfield  {author} {\bibinfo {author} {\bibfnamefont {G.}~\bibnamefont {Knizia}}\ and\ \bibinfo {author} {\bibfnamefont {G.~K.-L.}\ \bibnamefont {Chan}},\ }\bibfield  {title} {\bibinfo {title} {Density matrix embedding: A strong-coupling quantum embedding theory},\ }\href@noop {} {\bibfield  {journal} {\bibinfo  {journal} {Journal of chemical theory and computation}\ }\textbf {\bibinfo {volume} {9}},\ \bibinfo {pages} {1428} (\bibinfo {year} {2013})}\BibitemShut {NoStop}%
\bibitem [{\citenamefont {Libisch}\ \emph {et~al.}(2014)\citenamefont {Libisch}, \citenamefont {Huang},\ and\ \citenamefont {Carter}}]{libisch2014embedded}%
  \BibitemOpen
  \bibfield  {author} {\bibinfo {author} {\bibfnamefont {F.}~\bibnamefont {Libisch}}, \bibinfo {author} {\bibfnamefont {C.}~\bibnamefont {Huang}},\ and\ \bibinfo {author} {\bibfnamefont {E.~A.}\ \bibnamefont {Carter}},\ }\bibfield  {title} {\bibinfo {title} {Embedded correlated wavefunction schemes: Theory and applications},\ }\href@noop {} {\bibfield  {journal} {\bibinfo  {journal} {Accounts of chemical research}\ }\textbf {\bibinfo {volume} {47}},\ \bibinfo {pages} {2768} (\bibinfo {year} {2014})}\BibitemShut {NoStop}%
\bibitem [{\citenamefont {Wesolowski}\ \emph {et~al.}(2015)\citenamefont {Wesolowski}, \citenamefont {Shedge},\ and\ \citenamefont {Zhou}}]{wesolowski2015frozen}%
  \BibitemOpen
  \bibfield  {author} {\bibinfo {author} {\bibfnamefont {T.~A.}\ \bibnamefont {Wesolowski}}, \bibinfo {author} {\bibfnamefont {S.}~\bibnamefont {Shedge}},\ and\ \bibinfo {author} {\bibfnamefont {X.}~\bibnamefont {Zhou}},\ }\bibfield  {title} {\bibinfo {title} {Frozen-density embedding strategy for multilevel simulations of electronic structure},\ }\href@noop {} {\bibfield  {journal} {\bibinfo  {journal} {Chemical reviews}\ }\textbf {\bibinfo {volume} {115}},\ \bibinfo {pages} {5891} (\bibinfo {year} {2015})}\BibitemShut {NoStop}%
\bibitem [{\citenamefont {Govind}\ \emph {et~al.}(1998)\citenamefont {Govind}, \citenamefont {Wang}, \citenamefont {Da~Silva},\ and\ \citenamefont {Carter}}]{govind1998accurate}%
  \BibitemOpen
  \bibfield  {author} {\bibinfo {author} {\bibfnamefont {N.}~\bibnamefont {Govind}}, \bibinfo {author} {\bibfnamefont {Y.}~\bibnamefont {Wang}}, \bibinfo {author} {\bibfnamefont {A.}~\bibnamefont {Da~Silva}},\ and\ \bibinfo {author} {\bibfnamefont {E.}~\bibnamefont {Carter}},\ }\bibfield  {title} {\bibinfo {title} {Accurate ab initio energetics of extended systems via explicit correlation embedded in a density functional environment},\ }\href@noop {} {\bibfield  {journal} {\bibinfo  {journal} {Chemical physics letters}\ }\textbf {\bibinfo {volume} {295}},\ \bibinfo {pages} {129} (\bibinfo {year} {1998})}\BibitemShut {NoStop}%
\bibitem [{\citenamefont {Fornace}\ \emph {et~al.}(2015)\citenamefont {Fornace}, \citenamefont {Lee}, \citenamefont {Miyamoto}, \citenamefont {Manby},\ and\ \citenamefont {Miller~III}}]{fornace2015embedded}%
  \BibitemOpen
  \bibfield  {author} {\bibinfo {author} {\bibfnamefont {M.~E.}\ \bibnamefont {Fornace}}, \bibinfo {author} {\bibfnamefont {J.}~\bibnamefont {Lee}}, \bibinfo {author} {\bibfnamefont {K.}~\bibnamefont {Miyamoto}}, \bibinfo {author} {\bibfnamefont {F.~R.}\ \bibnamefont {Manby}},\ and\ \bibinfo {author} {\bibfnamefont {T.~F.}\ \bibnamefont {Miller~III}},\ }\bibfield  {title} {\bibinfo {title} {Embedded mean-field theory},\ }\href@noop {} {\bibfield  {journal} {\bibinfo  {journal} {Journal of chemical theory and computation}\ }\textbf {\bibinfo {volume} {11}},\ \bibinfo {pages} {568} (\bibinfo {year} {2015})}\BibitemShut {NoStop}%
\bibitem [{\citenamefont {H{\'e}gely}\ \emph {et~al.}(2016)\citenamefont {H{\'e}gely}, \citenamefont {Nagy}, \citenamefont {Ferenczy},\ and\ \citenamefont {K{\'a}llay}}]{hegely2016exact}%
  \BibitemOpen
  \bibfield  {author} {\bibinfo {author} {\bibfnamefont {B.}~\bibnamefont {H{\'e}gely}}, \bibinfo {author} {\bibfnamefont {P.~R.}\ \bibnamefont {Nagy}}, \bibinfo {author} {\bibfnamefont {G.~G.}\ \bibnamefont {Ferenczy}},\ and\ \bibinfo {author} {\bibfnamefont {M.}~\bibnamefont {K{\'a}llay}},\ }\bibfield  {title} {\bibinfo {title} {Exact density functional and wave function embedding schemes based on orbital localization},\ }\href@noop {} {\bibfield  {journal} {\bibinfo  {journal} {The Journal of Chemical Physics}\ }\textbf {\bibinfo {volume} {145}} (\bibinfo {year} {2016})}\BibitemShut {NoStop}%
\bibitem [{\citenamefont {Kowalski}(2018)}]{kowalski2018properties}%
  \BibitemOpen
  \bibfield  {author} {\bibinfo {author} {\bibfnamefont {K.}~\bibnamefont {Kowalski}},\ }\bibfield  {title} {\bibinfo {title} {Properties of coupled-cluster equations originating in excitation sub-algebras},\ }\href@noop {} {\bibfield  {journal} {\bibinfo  {journal} {The Journal of Chemical Physics}\ }\textbf {\bibinfo {volume} {148}} (\bibinfo {year} {2018})}\BibitemShut {NoStop}%
\bibitem [{\citenamefont {Bauman}\ \emph {et~al.}(2019)\citenamefont {Bauman}, \citenamefont {Bylaska}, \citenamefont {Krishnamoorthy}, \citenamefont {Low}, \citenamefont {Wiebe}, \citenamefont {Granade}, \citenamefont {Roetteler}, \citenamefont {Troyer},\ and\ \citenamefont {Kowalski}}]{bauman2019downfolding}%
  \BibitemOpen
  \bibfield  {author} {\bibinfo {author} {\bibfnamefont {N.~P.}\ \bibnamefont {Bauman}}, \bibinfo {author} {\bibfnamefont {E.~J.}\ \bibnamefont {Bylaska}}, \bibinfo {author} {\bibfnamefont {S.}~\bibnamefont {Krishnamoorthy}}, \bibinfo {author} {\bibfnamefont {G.~H.}\ \bibnamefont {Low}}, \bibinfo {author} {\bibfnamefont {N.}~\bibnamefont {Wiebe}}, \bibinfo {author} {\bibfnamefont {C.~E.}\ \bibnamefont {Granade}}, \bibinfo {author} {\bibfnamefont {M.}~\bibnamefont {Roetteler}}, \bibinfo {author} {\bibfnamefont {M.}~\bibnamefont {Troyer}},\ and\ \bibinfo {author} {\bibfnamefont {K.}~\bibnamefont {Kowalski}},\ }\bibfield  {title} {\bibinfo {title} {Downfolding of many-body hamiltonians using active-space models: Extension of the sub-system embedding sub-algebras approach to unitary coupled cluster formalisms},\ }\href@noop {} {\bibfield  {journal} {\bibinfo  {journal} {The Journal of chemical physics}\ }\textbf {\bibinfo {volume} {151}} (\bibinfo {year} {2019})}\BibitemShut {NoStop}%
\bibitem [{\citenamefont {Kowalski}\ and\ \citenamefont {Bauman}(2020)}]{kowalski2020sub}%
  \BibitemOpen
  \bibfield  {author} {\bibinfo {author} {\bibfnamefont {K.}~\bibnamefont {Kowalski}}\ and\ \bibinfo {author} {\bibfnamefont {N.~P.}\ \bibnamefont {Bauman}},\ }\bibfield  {title} {\bibinfo {title} {Sub-system quantum dynamics using coupled cluster downfolding techniques},\ }\href@noop {} {\bibfield  {journal} {\bibinfo  {journal} {The Journal of chemical physics}\ }\textbf {\bibinfo {volume} {152}} (\bibinfo {year} {2020})}\BibitemShut {NoStop}%
\bibitem [{\citenamefont {Kowalski}(2021)}]{kowalski2021dimensionality}%
  \BibitemOpen
  \bibfield  {author} {\bibinfo {author} {\bibfnamefont {K.}~\bibnamefont {Kowalski}},\ }\bibfield  {title} {\bibinfo {title} {Dimensionality reduction of the many-body problem using coupled-cluster subsystem flow equations: Classical and quantum computing perspective},\ }\href@noop {} {\bibfield  {journal} {\bibinfo  {journal} {Physical Review A}\ }\textbf {\bibinfo {volume} {104}},\ \bibinfo {pages} {032804} (\bibinfo {year} {2021})}\BibitemShut {NoStop}%
\bibitem [{\citenamefont {Bauman}\ and\ \citenamefont {Kowalski}(2022)}]{bauman2022coupled}%
  \BibitemOpen
  \bibfield  {author} {\bibinfo {author} {\bibfnamefont {N.~P.}\ \bibnamefont {Bauman}}\ and\ \bibinfo {author} {\bibfnamefont {K.}~\bibnamefont {Kowalski}},\ }\bibfield  {title} {\bibinfo {title} {Coupled cluster downfolding theory: Towards universal many-body algorithms for dimensionality reduction of composite quantum systems in chemistry and materials science},\ }\href@noop {} {\bibfield  {journal} {\bibinfo  {journal} {Materials Theory}\ }\textbf {\bibinfo {volume} {6}},\ \bibinfo {pages} {17} (\bibinfo {year} {2022})}\BibitemShut {NoStop}%
\bibitem [{\citenamefont {Kowalski}\ and\ \citenamefont {Bauman}(2023)}]{kowalski2023quantum}%
  \BibitemOpen
  \bibfield  {author} {\bibinfo {author} {\bibfnamefont {K.}~\bibnamefont {Kowalski}}\ and\ \bibinfo {author} {\bibfnamefont {N.~P.}\ \bibnamefont {Bauman}},\ }\bibfield  {title} {\bibinfo {title} {Quantum flow algorithms for simulating many-body systems on quantum computers},\ }\href@noop {} {\bibfield  {journal} {\bibinfo  {journal} {Physical Review Letters}\ }\textbf {\bibinfo {volume} {131}},\ \bibinfo {pages} {200601} (\bibinfo {year} {2023})}\BibitemShut {NoStop}%
\bibitem [{\citenamefont {Coester}(1958)}]{coester58_421}%
  \BibitemOpen
  \bibfield  {author} {\bibinfo {author} {\bibfnamefont {F.}~\bibnamefont {Coester}},\ }\bibfield  {title} {\bibinfo {title} {Bound states of a many-particle system},\ }\href {https://doi.org/http://dx.doi.org/10.1016/0029-5582(58)90280-3} {\bibfield  {journal} {\bibinfo  {journal} {Nucl. Phys.}\ }\textbf {\bibinfo {volume} {7}},\ \bibinfo {pages} {421} (\bibinfo {year} {1958})}\BibitemShut {NoStop}%
\bibitem [{\citenamefont {Coester}\ and\ \citenamefont {Kummel}(1960)}]{coester60_477}%
  \BibitemOpen
  \bibfield  {author} {\bibinfo {author} {\bibfnamefont {F.}~\bibnamefont {Coester}}\ and\ \bibinfo {author} {\bibfnamefont {H.}~\bibnamefont {Kummel}},\ }\bibfield  {title} {\bibinfo {title} {Short-range correlations in nuclear wave functions},\ }\href {https://doi.org/http://dx.doi.org/10.1016/0029-5582(60)90140-1} {\bibfield  {journal} {\bibinfo  {journal} {Nucl. Phys.}\ }\textbf {\bibinfo {volume} {17}},\ \bibinfo {pages} {477} (\bibinfo {year} {1960})}\BibitemShut {NoStop}%
\bibitem [{\citenamefont {{\v C}{\'\i}{\v z}ek}(1966)}]{cizek66_4256}%
  \BibitemOpen
  \bibfield  {author} {\bibinfo {author} {\bibfnamefont {J.}~\bibnamefont {{\v C}{\'\i}{\v z}ek}},\ }\bibfield  {title} {\bibinfo {title} {On the correlation problem in atomic and molecular systems. calculation of wavefunction components in ursell-type expansion using quantum-field theoretical methods},\ }\href {https://doi.org/http://dx.doi.org/10.1063/1.1727484} {\bibfield  {journal} {\bibinfo  {journal} {The Journal of Chemical Physics}\ }\textbf {\bibinfo {volume} {45}},\ \bibinfo {pages} {4256} (\bibinfo {year} {1966})}\BibitemShut {NoStop}%
\bibitem [{\citenamefont {Paldus}\ \emph {et~al.}(1972)\citenamefont {Paldus}, \citenamefont {{\v{C}}{\'\i}{\v{z}}ek},\ and\ \citenamefont {Shavitt}}]{paldus1972correlation}%
  \BibitemOpen
  \bibfield  {author} {\bibinfo {author} {\bibfnamefont {J.}~\bibnamefont {Paldus}}, \bibinfo {author} {\bibfnamefont {J.}~\bibnamefont {{\v{C}}{\'\i}{\v{z}}ek}},\ and\ \bibinfo {author} {\bibfnamefont {I.}~\bibnamefont {Shavitt}},\ }\bibfield  {title} {\bibinfo {title} {Correlation problems in atomic and molecular systems. iv. extended coupled-pair many-electron theory and its application to the bh$_3$ molecule},\ }\href@noop {} {\bibfield  {journal} {\bibinfo  {journal} {Phys. Rev. A}\ }\textbf {\bibinfo {volume} {5}},\ \bibinfo {pages} {50} (\bibinfo {year} {1972})}\BibitemShut {NoStop}%
\bibitem [{\citenamefont {Purvis}\ and\ \citenamefont {Bartlett}(1982)}]{purvis82_1910}%
  \BibitemOpen
  \bibfield  {author} {\bibinfo {author} {\bibfnamefont {G.~D.}\ \bibnamefont {Purvis}}\ and\ \bibinfo {author} {\bibfnamefont {R.~J.}\ \bibnamefont {Bartlett}},\ }\bibfield  {title} {\bibinfo {title} {A full coupled-cluster singles and doubles model: The inclusion of disconnected triples},\ }\href {https://doi.org/http://dx.doi.org/10.1063/1.443164} {\bibfield  {journal} {\bibinfo  {journal} {The Journal of Chemical Physics}\ }\textbf {\bibinfo {volume} {76}},\ \bibinfo {pages} {1910} (\bibinfo {year} {1982})}\BibitemShut {NoStop}%
\bibitem [{\citenamefont {Raghavachari}\ \emph {et~al.}(1989)\citenamefont {Raghavachari}, \citenamefont {Trucks}, \citenamefont {Pople},\ and\ \citenamefont {Head-Gordon}}]{raghavachari89_479}%
  \BibitemOpen
  \bibfield  {author} {\bibinfo {author} {\bibfnamefont {K.}~\bibnamefont {Raghavachari}}, \bibinfo {author} {\bibfnamefont {G.~W.}\ \bibnamefont {Trucks}}, \bibinfo {author} {\bibfnamefont {J.~A.}\ \bibnamefont {Pople}},\ and\ \bibinfo {author} {\bibfnamefont {M.}~\bibnamefont {Head-Gordon}},\ }\bibfield  {title} {\bibinfo {title} {A fifth-order perturbation comparison of electron correlation theories},\ }\href {https://doi.org/http://dx.doi.org/10.1016/S0009-2614(89)87395-6} {\bibfield  {journal} {\bibinfo  {journal} {Chem. Phys. Lett.}\ }\textbf {\bibinfo {volume} {157}},\ \bibinfo {pages} {479} (\bibinfo {year} {1989})}\BibitemShut {NoStop}%
\bibitem [{\citenamefont {Paldus}\ and\ \citenamefont {Li}(2007)}]{paldus07}%
  \BibitemOpen
  \bibfield  {author} {\bibinfo {author} {\bibfnamefont {J.}~\bibnamefont {Paldus}}\ and\ \bibinfo {author} {\bibfnamefont {X.}~\bibnamefont {Li}},\ }\bibinfo {title} {A critical assessment of coupled cluster method in quantum chemistry},\ in\ \href {https://doi.org/10.1002/9780470141694.ch1} {\emph {\bibinfo {booktitle} {Advances in Chemical Physics}}}\ (\bibinfo  {publisher} {John Wiley \& Sons, Ltd},\ \bibinfo {year} {2007})\ pp.\ \bibinfo {pages} {1--175}\BibitemShut {NoStop}%
\bibitem [{\citenamefont {Crawford}\ and\ \citenamefont {Schaefer}(2000)}]{crawford2000introduction}%
  \BibitemOpen
  \bibfield  {author} {\bibinfo {author} {\bibfnamefont {T.~D.}\ \bibnamefont {Crawford}}\ and\ \bibinfo {author} {\bibfnamefont {H.~F.}\ \bibnamefont {Schaefer}},\ }\bibfield  {title} {\bibinfo {title} {{An Introduction to Coupled Cluster Theory for Computational Chemists}},\ }\href {https://doi.org/10.1002/9780470125915.ch2} {\bibfield  {journal} {\bibinfo  {journal} {Reviews in Computational Chemistry}\ }\textbf {\bibinfo {volume} {14}},\ \bibinfo {pages} {33} (\bibinfo {year} {2000})}\BibitemShut {NoStop}%
\bibitem [{\citenamefont {Bartlett}\ and\ \citenamefont {Musia\l}(2007)}]{bartlett07_291}%
  \BibitemOpen
  \bibfield  {author} {\bibinfo {author} {\bibfnamefont {R.~J.}\ \bibnamefont {Bartlett}}\ and\ \bibinfo {author} {\bibfnamefont {M.}~\bibnamefont {Musia\l}},\ }\bibfield  {title} {\bibinfo {title} {{Coupled-Cluster Theory in Quantum Chemistry}},\ }\href {https://doi.org/10.1103/RevModPhys.79.291} {\bibfield  {journal} {\bibinfo  {journal} {Rev. Mod. Phys.}\ }\textbf {\bibinfo {volume} {79}},\ \bibinfo {pages} {291} (\bibinfo {year} {2007})}\BibitemShut {NoStop}%
\bibitem [{\citenamefont {Kowalski}(2023)}]{kowalski2023sub}%
  \BibitemOpen
  \bibfield  {author} {\bibinfo {author} {\bibfnamefont {K.}~\bibnamefont {Kowalski}},\ }\bibfield  {title} {\bibinfo {title} {Sub-system self-consistency in coupled cluster theory},\ }\href@noop {} {\bibfield  {journal} {\bibinfo  {journal} {The Journal of Chemical Physics}\ }\textbf {\bibinfo {volume} {158}} (\bibinfo {year} {2023})}\BibitemShut {NoStop}%
\bibitem [{\citenamefont {Peng}\ and\ \citenamefont {Kowalski}(2024)}]{peng2024integrating}%
  \BibitemOpen
  \bibfield  {author} {\bibinfo {author} {\bibfnamefont {B.}~\bibnamefont {Peng}}\ and\ \bibinfo {author} {\bibfnamefont {K.}~\bibnamefont {Kowalski}},\ }\bibfield  {title} {\bibinfo {title} {Integrating subsystem embedding subalgebras and coupled cluster green’s function: a theoretical foundation for quantum embedding in excitation manifold},\ }\href@noop {} {\bibfield  {journal} {\bibinfo  {journal} {Electronic Structure}\ }\textbf {\bibinfo {volume} {6}},\ \bibinfo {pages} {015005} (\bibinfo {year} {2024})}\BibitemShut {NoStop}%
\bibitem [{\citenamefont {Shee}\ \emph {et~al.}(2024)\citenamefont {Shee}, \citenamefont {Faulstich}, \citenamefont {Whaley}, \citenamefont {Lin},\ and\ \citenamefont {Head-Gordon}}]{shee2024static}%
  \BibitemOpen
  \bibfield  {author} {\bibinfo {author} {\bibfnamefont {A.}~\bibnamefont {Shee}}, \bibinfo {author} {\bibfnamefont {F.~M.}\ \bibnamefont {Faulstich}}, \bibinfo {author} {\bibfnamefont {K.~B.}\ \bibnamefont {Whaley}}, \bibinfo {author} {\bibfnamefont {L.}~\bibnamefont {Lin}},\ and\ \bibinfo {author} {\bibfnamefont {M.}~\bibnamefont {Head-Gordon}},\ }\bibfield  {title} {\bibinfo {title} {A static quantum embedding scheme based on coupled cluster theory},\ }\href@noop {} {\bibfield  {journal} {\bibinfo  {journal} {The Journal of Chemical Physics}\ }\textbf {\bibinfo {volume} {161}} (\bibinfo {year} {2024})}\BibitemShut {NoStop}%
\bibitem [{\citenamefont {Liang}\ \emph {et~al.}(2024)\citenamefont {Liang}, \citenamefont {Kowalski}, \citenamefont {Yang},\ and\ \citenamefont {Bauman}}]{liang2024effective}%
  \BibitemOpen
  \bibfield  {author} {\bibinfo {author} {\bibfnamefont {S.}~\bibnamefont {Liang}}, \bibinfo {author} {\bibfnamefont {K.}~\bibnamefont {Kowalski}}, \bibinfo {author} {\bibfnamefont {C.}~\bibnamefont {Yang}},\ and\ \bibinfo {author} {\bibfnamefont {N.~P.}\ \bibnamefont {Bauman}},\ }\bibfield  {title} {\bibinfo {title} {Effective many-body interactions in reduced-dimensionality spaces through neural network models},\ }\href@noop {} {\bibfield  {journal} {\bibinfo  {journal} {Physical Review Research}\ }\textbf {\bibinfo {volume} {6}},\ \bibinfo {pages} {043287} (\bibinfo {year} {2024})}\BibitemShut {NoStop}%
\bibitem [{\citenamefont {Bylaska}\ \emph {et~al.}(2024)\citenamefont {Bylaska}, \citenamefont {Panyala}, \citenamefont {Bauman}, \citenamefont {Peng}, \citenamefont {Pathak}, \citenamefont {Mejia-Rodriguez}, \citenamefont {Govind}, \citenamefont {Williams-Young}, \citenamefont {Apr{\`a}}, \citenamefont {Bagusetty} \emph {et~al.}}]{bylaska2024electronic}%
  \BibitemOpen
  \bibfield  {author} {\bibinfo {author} {\bibfnamefont {E.~J.}\ \bibnamefont {Bylaska}}, \bibinfo {author} {\bibfnamefont {A.}~\bibnamefont {Panyala}}, \bibinfo {author} {\bibfnamefont {N.~P.}\ \bibnamefont {Bauman}}, \bibinfo {author} {\bibfnamefont {B.}~\bibnamefont {Peng}}, \bibinfo {author} {\bibfnamefont {H.}~\bibnamefont {Pathak}}, \bibinfo {author} {\bibfnamefont {D.}~\bibnamefont {Mejia-Rodriguez}}, \bibinfo {author} {\bibfnamefont {N.}~\bibnamefont {Govind}}, \bibinfo {author} {\bibfnamefont {D.~B.}\ \bibnamefont {Williams-Young}}, \bibinfo {author} {\bibfnamefont {E.}~\bibnamefont {Apr{\`a}}}, \bibinfo {author} {\bibfnamefont {A.}~\bibnamefont {Bagusetty}}, \emph {et~al.},\ }\bibfield  {title} {\bibinfo {title} {Electronic structure simulations in the cloud computing environment},\ }\href@noop {} {\bibfield  {journal} {\bibinfo  {journal} {The Journal of Chemical Physics}\ }\textbf {\bibinfo {volume} {161}} (\bibinfo {year} {2024})}\BibitemShut {NoStop}%
\bibitem [{\citenamefont {Mutlu}\ \emph {et~al.}(2023)\citenamefont {Mutlu}, \citenamefont {Panyala}, \citenamefont {Gawande}, \citenamefont {Bagusetty}, \citenamefont {Glabe}, \citenamefont {Kim}, \citenamefont {Kowalski}, \citenamefont {Bauman}, \citenamefont {Peng}, \citenamefont {Pathak} \emph {et~al.}}]{mutlu2023tamm}%
  \BibitemOpen
  \bibfield  {author} {\bibinfo {author} {\bibfnamefont {E.}~\bibnamefont {Mutlu}}, \bibinfo {author} {\bibfnamefont {A.}~\bibnamefont {Panyala}}, \bibinfo {author} {\bibfnamefont {N.}~\bibnamefont {Gawande}}, \bibinfo {author} {\bibfnamefont {A.}~\bibnamefont {Bagusetty}}, \bibinfo {author} {\bibfnamefont {J.}~\bibnamefont {Glabe}}, \bibinfo {author} {\bibfnamefont {J.}~\bibnamefont {Kim}}, \bibinfo {author} {\bibfnamefont {K.}~\bibnamefont {Kowalski}}, \bibinfo {author} {\bibfnamefont {N.~P.}\ \bibnamefont {Bauman}}, \bibinfo {author} {\bibfnamefont {B.}~\bibnamefont {Peng}}, \bibinfo {author} {\bibfnamefont {H.}~\bibnamefont {Pathak}}, \emph {et~al.},\ }\bibfield  {title} {\bibinfo {title} {Tamm: Tensor algebra for many-body methods},\ }\href@noop {} {\bibfield  {journal} {\bibinfo  {journal} {The Journal of Chemical Physics}\ }\textbf {\bibinfo {volume} {159}} (\bibinfo {year} {2023})}\BibitemShut {NoStop}%
\bibitem [{\citenamefont {Oliphant}\ and\ \citenamefont {Adamowicz}(1992)}]{oliphant1992implementation}%
  \BibitemOpen
  \bibfield  {author} {\bibinfo {author} {\bibfnamefont {N.}~\bibnamefont {Oliphant}}\ and\ \bibinfo {author} {\bibfnamefont {L.}~\bibnamefont {Adamowicz}},\ }\bibfield  {title} {\bibinfo {title} {The implementation of the multireference coupled-cluster method based on the single-reference formalism},\ }\href@noop {} {\bibfield  {journal} {\bibinfo  {journal} {The Journal of chemical physics}\ }\textbf {\bibinfo {volume} {96}},\ \bibinfo {pages} {3739} (\bibinfo {year} {1992})}\BibitemShut {NoStop}%
\bibitem [{\citenamefont {Oliphant}\ and\ \citenamefont {Adamowicz}(1993)}]{oliphant1993multireference}%
  \BibitemOpen
  \bibfield  {author} {\bibinfo {author} {\bibfnamefont {N.}~\bibnamefont {Oliphant}}\ and\ \bibinfo {author} {\bibfnamefont {L.}~\bibnamefont {Adamowicz}},\ }\bibfield  {title} {\bibinfo {title} {Multireference coupled cluster method for electronic structure of molecules},\ }\href@noop {} {\bibfield  {journal} {\bibinfo  {journal} {International Reviews in Physical Chemistry}\ }\textbf {\bibinfo {volume} {12}},\ \bibinfo {pages} {339} (\bibinfo {year} {1993})}\BibitemShut {NoStop}%
\bibitem [{\citenamefont {Piecuch}\ \emph {et~al.}(1993)\citenamefont {Piecuch}, \citenamefont {Oliphant},\ and\ \citenamefont {Adamowicz}}]{pnl93}%
  \BibitemOpen
  \bibfield  {author} {\bibinfo {author} {\bibfnamefont {P.}~\bibnamefont {Piecuch}}, \bibinfo {author} {\bibfnamefont {N.}~\bibnamefont {Oliphant}},\ and\ \bibinfo {author} {\bibfnamefont {L.}~\bibnamefont {Adamowicz}},\ }\bibfield  {title} {\bibinfo {title} {A state-selective multireference coupled-cluster theory employing the single-reference formalism},\ }\href {https://doi.org/10.1063/1.466179} {\bibfield  {journal} {\bibinfo  {journal} {J. Chem. Phys.}\ }\textbf {\bibinfo {volume} {99}},\ \bibinfo {pages} {1875} (\bibinfo {year} {1993})}\BibitemShut {NoStop}%
\bibitem [{\citenamefont {Adamowicz}\ \emph {et~al.}(1998)\citenamefont {Adamowicz}, \citenamefont {Piecuch},\ and\ \citenamefont {Ghose}}]{adamowicz1998state}%
  \BibitemOpen
  \bibfield  {author} {\bibinfo {author} {\bibfnamefont {L.}~\bibnamefont {Adamowicz}}, \bibinfo {author} {\bibfnamefont {P.}~\bibnamefont {Piecuch}},\ and\ \bibinfo {author} {\bibfnamefont {K.~B.}\ \bibnamefont {Ghose}},\ }\bibfield  {title} {\bibinfo {title} {The state-selective coupled cluster method for quasi-degenerate electronic states},\ }\href@noop {} {\bibfield  {journal} {\bibinfo  {journal} {Molecular Physics}\ }\textbf {\bibinfo {volume} {94}},\ \bibinfo {pages} {225} (\bibinfo {year} {1998})}\BibitemShut {NoStop}%
\bibitem [{\citenamefont {Elfwing}\ \emph {et~al.}(2018)\citenamefont {Elfwing}, \citenamefont {Uchibe},\ and\ \citenamefont {Doya}}]{elfwing2018sigmoid}%
  \BibitemOpen
  \bibfield  {author} {\bibinfo {author} {\bibfnamefont {S.}~\bibnamefont {Elfwing}}, \bibinfo {author} {\bibfnamefont {E.}~\bibnamefont {Uchibe}},\ and\ \bibinfo {author} {\bibfnamefont {K.}~\bibnamefont {Doya}},\ }\bibfield  {title} {\bibinfo {title} {Sigmoid-weighted linear units for neural network function approximation in reinforcement learning},\ }\href@noop {} {\bibfield  {journal} {\bibinfo  {journal} {Neural networks}\ }\textbf {\bibinfo {volume} {107}},\ \bibinfo {pages} {3} (\bibinfo {year} {2018})}\BibitemShut {NoStop}%
\bibitem [{\citenamefont {Van~Vleck}(1929)}]{van1929sigma}%
  \BibitemOpen
  \bibfield  {author} {\bibinfo {author} {\bibfnamefont {J.}~\bibnamefont {Van~Vleck}},\ }\bibfield  {title} {\bibinfo {title} {On $\sigma$-type doubling and electron spin in the spectra of diatomic molecules},\ }\href@noop {} {\bibfield  {journal} {\bibinfo  {journal} {Physical Review}\ }\textbf {\bibinfo {volume} {33}},\ \bibinfo {pages} {467} (\bibinfo {year} {1929})}\BibitemShut {NoStop}%
\bibitem [{\citenamefont {Jordahl}(1934)}]{jordahl1934effect}%
  \BibitemOpen
  \bibfield  {author} {\bibinfo {author} {\bibfnamefont {O.}~\bibnamefont {Jordahl}},\ }\bibfield  {title} {\bibinfo {title} {The effect of crystalline electric fields on the paramagnetic susceptibility of cupric salts},\ }\href@noop {} {\bibfield  {journal} {\bibinfo  {journal} {Physical Review}\ }\textbf {\bibinfo {volume} {45}},\ \bibinfo {pages} {87} (\bibinfo {year} {1934})}\BibitemShut {NoStop}%
\bibitem [{\citenamefont {Bloch}(1958)}]{bloch1958theorie}%
  \BibitemOpen
  \bibfield  {author} {\bibinfo {author} {\bibfnamefont {C.}~\bibnamefont {Bloch}},\ }\bibfield  {title} {\bibinfo {title} {Sur la th{\'e}orie des perturbations des {\'e}tats li{\'e}s},\ }\href@noop {} {\bibfield  {journal} {\bibinfo  {journal} {Nuclear Physics}\ }\textbf {\bibinfo {volume} {6}},\ \bibinfo {pages} {329} (\bibinfo {year} {1958})}\BibitemShut {NoStop}%
\bibitem [{\citenamefont {Des~Cloizeaux}(1960)}]{des1960extension}%
  \BibitemOpen
  \bibfield  {author} {\bibinfo {author} {\bibfnamefont {J.}~\bibnamefont {Des~Cloizeaux}},\ }\bibfield  {title} {\bibinfo {title} {Extension d'une formule de lagrange {\`a} des probl{\`e}mes de valeurs propres},\ }\href@noop {} {\bibfield  {journal} {\bibinfo  {journal} {Nuclear Physics}\ }\textbf {\bibinfo {volume} {20}},\ \bibinfo {pages} {321} (\bibinfo {year} {1960})}\BibitemShut {NoStop}%
\bibitem [{\citenamefont {Primas}(1961)}]{primas1961verallgemeinerte}%
  \BibitemOpen
  \bibfield  {author} {\bibinfo {author} {\bibfnamefont {H.}~\bibnamefont {Primas}},\ }\bibfield  {title} {\bibinfo {title} {Eine verallgemeinerte st{\"o}rungstheorie f{\"u}r quantenmechanische mehrteilchenprobleme},\ }\href@noop {} {\bibfield  {journal} {\bibinfo  {journal} {Helv. Phys. Acta}\ }\textbf {\bibinfo {volume} {34}},\ \bibinfo {pages} {331} (\bibinfo {year} {1961})}\BibitemShut {NoStop}%
\bibitem [{\citenamefont {Primas}(1963)}]{primas1963generalized}%
  \BibitemOpen
  \bibfield  {author} {\bibinfo {author} {\bibfnamefont {H.}~\bibnamefont {Primas}},\ }\bibfield  {title} {\bibinfo {title} {Generalized perturbation theory in operator form},\ }\href@noop {} {\bibfield  {journal} {\bibinfo  {journal} {Reviews of Modern Physics}\ }\textbf {\bibinfo {volume} {35}},\ \bibinfo {pages} {710} (\bibinfo {year} {1963})}\BibitemShut {NoStop}%
\bibitem [{\citenamefont {L{\"o}wdin}(1963)}]{lowdin1963studies}%
  \BibitemOpen
  \bibfield  {author} {\bibinfo {author} {\bibfnamefont {P.-O.}\ \bibnamefont {L{\"o}wdin}},\ }\bibfield  {title} {\bibinfo {title} {Studies in perturbation theory: Part i. an elementary iteration-variation procedure for solving the schr{\"o}dinger equation by partitioning technique},\ }\href@noop {} {\bibfield  {journal} {\bibinfo  {journal} {Journal of Molecular Spectroscopy}\ }\textbf {\bibinfo {volume} {10}},\ \bibinfo {pages} {12} (\bibinfo {year} {1963})}\BibitemShut {NoStop}%
\bibitem [{\citenamefont {Schrieffer}\ and\ \citenamefont {Wolff}(1966)}]{schrieffer1966relation}%
  \BibitemOpen
  \bibfield  {author} {\bibinfo {author} {\bibfnamefont {J.~R.}\ \bibnamefont {Schrieffer}}\ and\ \bibinfo {author} {\bibfnamefont {P.~A.}\ \bibnamefont {Wolff}},\ }\bibfield  {title} {\bibinfo {title} {Relation between the anderson and kondo hamiltonians},\ }\href@noop {} {\bibfield  {journal} {\bibinfo  {journal} {Physical Review}\ }\textbf {\bibinfo {volume} {149}},\ \bibinfo {pages} {491} (\bibinfo {year} {1966})}\BibitemShut {NoStop}%
\bibitem [{\citenamefont {Kirtman}(1968)}]{kirtman1968variational}%
  \BibitemOpen
  \bibfield  {author} {\bibinfo {author} {\bibfnamefont {B.}~\bibnamefont {Kirtman}},\ }\bibfield  {title} {\bibinfo {title} {Variational form of van vleck degenerate perturbation theory with particular application to electronic structure problems},\ }\href@noop {} {\bibfield  {journal} {\bibinfo  {journal} {Journal of Chemical Physics}\ }\textbf {\bibinfo {volume} {49}},\ \bibinfo {pages} {3890} (\bibinfo {year} {1968})}\BibitemShut {NoStop}%
\bibitem [{\citenamefont {Soliverez}(1969)}]{soliverez1969effective}%
  \BibitemOpen
  \bibfield  {author} {\bibinfo {author} {\bibfnamefont {C.}~\bibnamefont {Soliverez}},\ }\bibfield  {title} {\bibinfo {title} {An effective hamiltonian and time-independent perturbation theory},\ }\href@noop {} {\bibfield  {journal} {\bibinfo  {journal} {Journal of Physics C: Solid State Physics}\ }\textbf {\bibinfo {volume} {2}},\ \bibinfo {pages} {2161} (\bibinfo {year} {1969})}\BibitemShut {NoStop}%
\bibitem [{\citenamefont {J{\o}rgensen}\ and\ \citenamefont {Pedersen}(1974)}]{jorgensen1974projector}%
  \BibitemOpen
  \bibfield  {author} {\bibinfo {author} {\bibfnamefont {F.}~\bibnamefont {J{\o}rgensen}}\ and\ \bibinfo {author} {\bibfnamefont {T.}~\bibnamefont {Pedersen}},\ }\bibfield  {title} {\bibinfo {title} {A projector formulation for the van vleck transformation: Ii. near-degenerate case},\ }\href@noop {} {\bibfield  {journal} {\bibinfo  {journal} {Molecular Physics}\ }\textbf {\bibinfo {volume} {27}},\ \bibinfo {pages} {959} (\bibinfo {year} {1974})}\BibitemShut {NoStop}%
\bibitem [{\citenamefont {J{\o}rgensen}(1975)}]{jorgensen1975effective}%
  \BibitemOpen
  \bibfield  {author} {\bibinfo {author} {\bibfnamefont {F.}~\bibnamefont {J{\o}rgensen}},\ }\bibfield  {title} {\bibinfo {title} {Effective hamiltonians},\ }\href@noop {} {\bibfield  {journal} {\bibinfo  {journal} {Molecular Physics}\ }\textbf {\bibinfo {volume} {29}},\ \bibinfo {pages} {1137} (\bibinfo {year} {1975})}\BibitemShut {NoStop}%
\bibitem [{\citenamefont {Mukherjee}\ \emph {et~al.}(1977)\citenamefont {Mukherjee}, \citenamefont {Moitra},\ and\ \citenamefont {Mukhopadhyay}}]{mukherjee1977ab}%
  \BibitemOpen
  \bibfield  {author} {\bibinfo {author} {\bibfnamefont {D.}~\bibnamefont {Mukherjee}}, \bibinfo {author} {\bibfnamefont {R.~K.}\ \bibnamefont {Moitra}},\ and\ \bibinfo {author} {\bibfnamefont {A.}~\bibnamefont {Mukhopadhyay}},\ }\bibfield  {title} {\bibinfo {title} {An ab-initio derivation of the pi-electron hamiltonian by a nonperturbative open-shell formalism},\ }\href@noop {} {\bibfield  {journal} {\bibinfo  {journal} {Pramana}\ }\textbf {\bibinfo {volume} {9}},\ \bibinfo {pages} {545} (\bibinfo {year} {1977})}\BibitemShut {NoStop}%
\bibitem [{\citenamefont {Lindgren}(1978)}]{lindgren1978coupled}%
  \BibitemOpen
  \bibfield  {author} {\bibinfo {author} {\bibfnamefont {I.}~\bibnamefont {Lindgren}},\ }\bibfield  {title} {\bibinfo {title} {A coupled-cluster approach to the many-body perturbation theory for open-shell systems},\ }\href@noop {} {\bibfield  {journal} {\bibinfo  {journal} {International Journal of Quantum Chemistry}\ }\textbf {\bibinfo {volume} {14}},\ \bibinfo {pages} {33} (\bibinfo {year} {1978})}\BibitemShut {NoStop}%
\bibitem [{\citenamefont {Lindgren}\ and\ \citenamefont {Morrison}(2012)}]{lindgren2012atomic}%
  \BibitemOpen
  \bibfield  {author} {\bibinfo {author} {\bibfnamefont {I.}~\bibnamefont {Lindgren}}\ and\ \bibinfo {author} {\bibfnamefont {J.}~\bibnamefont {Morrison}},\ }\href@noop {} {\emph {\bibinfo {title} {Atomic many-body theory}}},\ Vol.~\bibinfo {volume} {3}\ (\bibinfo  {publisher} {Springer Science \& Business Media},\ \bibinfo {year} {2012})\BibitemShut {NoStop}%
\bibitem [{\citenamefont {Shavitt}\ and\ \citenamefont {Redmon}(1980)}]{shavitt1980quasidegenerate}%
  \BibitemOpen
  \bibfield  {author} {\bibinfo {author} {\bibfnamefont {I.}~\bibnamefont {Shavitt}}\ and\ \bibinfo {author} {\bibfnamefont {L.~T.}\ \bibnamefont {Redmon}},\ }\bibfield  {title} {\bibinfo {title} {Quasidegenerate perturbation theories. a canonical van vleck formalism and its relationship to other approaches},\ }\href@noop {} {\bibfield  {journal} {\bibinfo  {journal} {The Journal of Chemical Physics}\ }\textbf {\bibinfo {volume} {73}},\ \bibinfo {pages} {5711} (\bibinfo {year} {1980})}\BibitemShut {NoStop}%
\bibitem [{\citenamefont {Westhaus}(1981)}]{westhaus1981connections}%
  \BibitemOpen
  \bibfield  {author} {\bibinfo {author} {\bibfnamefont {P.}~\bibnamefont {Westhaus}},\ }\bibfield  {title} {\bibinfo {title} {Connections between perturbation theory and the unitary transformation methods of deriving effective hamiltonians},\ }\href@noop {} {\bibfield  {journal} {\bibinfo  {journal} {International Journal of Quantum Chemistry}\ }\textbf {\bibinfo {volume} {20}},\ \bibinfo {pages} {1243} (\bibinfo {year} {1981})}\BibitemShut {NoStop}%
\bibitem [{\citenamefont {Suzuki}(1982)}]{suzuki1982construction}%
  \BibitemOpen
  \bibfield  {author} {\bibinfo {author} {\bibfnamefont {K.}~\bibnamefont {Suzuki}},\ }\bibfield  {title} {\bibinfo {title} {Construction of hermitian effective interaction in nuclei: —general relation between hermitian and non-hermitian forms—},\ }\href@noop {} {\bibfield  {journal} {\bibinfo  {journal} {Progress of Theoretical Physics}\ }\textbf {\bibinfo {volume} {68}},\ \bibinfo {pages} {246} (\bibinfo {year} {1982})}\BibitemShut {NoStop}%
\bibitem [{\citenamefont {Suzuki}\ and\ \citenamefont {Okamoto}(1983)}]{suzuki1983degenerate}%
  \BibitemOpen
  \bibfield  {author} {\bibinfo {author} {\bibfnamefont {K.}~\bibnamefont {Suzuki}}\ and\ \bibinfo {author} {\bibfnamefont {R.}~\bibnamefont {Okamoto}},\ }\bibfield  {title} {\bibinfo {title} {Degenerate perturbation theory in quantum mechanics},\ }\href@noop {} {\bibfield  {journal} {\bibinfo  {journal} {Progress of Theoretical Physics}\ }\textbf {\bibinfo {volume} {70}},\ \bibinfo {pages} {439} (\bibinfo {year} {1983})}\BibitemShut {NoStop}%
\bibitem [{\citenamefont {G{\l}azek}\ and\ \citenamefont {Wilson}(1993)}]{glazek1993renormalization}%
  \BibitemOpen
  \bibfield  {author} {\bibinfo {author} {\bibfnamefont {S.~D.}\ \bibnamefont {G{\l}azek}}\ and\ \bibinfo {author} {\bibfnamefont {K.~G.}\ \bibnamefont {Wilson}},\ }\bibfield  {title} {\bibinfo {title} {Renormalization of hamiltonians},\ }\href@noop {} {\bibfield  {journal} {\bibinfo  {journal} {Physical Review D}\ }\textbf {\bibinfo {volume} {48}},\ \bibinfo {pages} {5863} (\bibinfo {year} {1993})}\BibitemShut {NoStop}%
\bibitem [{\citenamefont {Bravyi}\ \emph {et~al.}(2011)\citenamefont {Bravyi}, \citenamefont {DiVincenzo},\ and\ \citenamefont {Loss}}]{bravyi2011schrieffer}%
  \BibitemOpen
  \bibfield  {author} {\bibinfo {author} {\bibfnamefont {S.}~\bibnamefont {Bravyi}}, \bibinfo {author} {\bibfnamefont {D.~P.}\ \bibnamefont {DiVincenzo}},\ and\ \bibinfo {author} {\bibfnamefont {D.}~\bibnamefont {Loss}},\ }\bibfield  {title} {\bibinfo {title} {Schrieffer--wolff transformation for quantum many-body systems},\ }\href@noop {} {\bibfield  {journal} {\bibinfo  {journal} {Annals of physics}\ }\textbf {\bibinfo {volume} {326}},\ \bibinfo {pages} {2793} (\bibinfo {year} {2011})}\BibitemShut {NoStop}%
\bibitem [{\citenamefont {Stroberg}\ \emph {et~al.}(2019)\citenamefont {Stroberg}, \citenamefont {Hergert}, \citenamefont {Bogner},\ and\ \citenamefont {Holt}}]{stroberg2019nonempirical}%
  \BibitemOpen
  \bibfield  {author} {\bibinfo {author} {\bibfnamefont {S.~R.}\ \bibnamefont {Stroberg}}, \bibinfo {author} {\bibfnamefont {H.}~\bibnamefont {Hergert}}, \bibinfo {author} {\bibfnamefont {S.~K.}\ \bibnamefont {Bogner}},\ and\ \bibinfo {author} {\bibfnamefont {J.~D.}\ \bibnamefont {Holt}},\ }\bibfield  {title} {\bibinfo {title} {Nonempirical interactions for the nuclear shell model: an update},\ }\href@noop {} {\bibfield  {journal} {\bibinfo  {journal} {Annual Review of Nuclear and Particle Science}\ }\textbf {\bibinfo {volume} {69}},\ \bibinfo {pages} {307} (\bibinfo {year} {2019})}\BibitemShut {NoStop}%
\bibitem [{\citenamefont {Durand}(1983)}]{durand1983direct}%
  \BibitemOpen
  \bibfield  {author} {\bibinfo {author} {\bibfnamefont {P.}~\bibnamefont {Durand}},\ }\bibfield  {title} {\bibinfo {title} {Direct determination of effective hamiltonians by wave-operator methods. i. general formalism},\ }\href@noop {} {\bibfield  {journal} {\bibinfo  {journal} {Physical Review A}\ }\textbf {\bibinfo {volume} {28}},\ \bibinfo {pages} {3184} (\bibinfo {year} {1983})}\BibitemShut {NoStop}%
\bibitem [{\citenamefont {Durand}\ and\ \citenamefont {Malrieu}(1987)}]{durand1987effective}%
  \BibitemOpen
  \bibfield  {author} {\bibinfo {author} {\bibfnamefont {P.}~\bibnamefont {Durand}}\ and\ \bibinfo {author} {\bibfnamefont {J.-P.}\ \bibnamefont {Malrieu}},\ }\bibfield  {title} {\bibinfo {title} {Effective hamiltonians and pseudo-operators as tools for rigorous modelling},\ }\href@noop {} {\bibfield  {journal} {\bibinfo  {journal} {Advances in Chemical Physics: Ab Initio Methods in Quantum Chemistry Part I}\ }\textbf {\bibinfo {volume} {67}},\ \bibinfo {pages} {321} (\bibinfo {year} {1987})}\BibitemShut {NoStop}%
\bibitem [{\citenamefont {Jeziorski}\ and\ \citenamefont {Monkhorst}(1981)}]{jezmonk}%
  \BibitemOpen
  \bibfield  {author} {\bibinfo {author} {\bibfnamefont {B.}~\bibnamefont {Jeziorski}}\ and\ \bibinfo {author} {\bibfnamefont {H.~J.}\ \bibnamefont {Monkhorst}},\ }\bibfield  {title} {\bibinfo {title} {Coupled-cluster method for multideterminantal reference states},\ }\href {https://doi.org/10.1103/PhysRevA.24.1668} {\bibfield  {journal} {\bibinfo  {journal} {Phys. Rev. A}\ }\textbf {\bibinfo {volume} {24}},\ \bibinfo {pages} {1668} (\bibinfo {year} {1981})}\BibitemShut {NoStop}%
\bibitem [{\citenamefont {Pal}\ \emph {et~al.}(1988)\citenamefont {Pal}, \citenamefont {Rittby}, \citenamefont {Bartlett}, \citenamefont {Sinha},\ and\ \citenamefont {Mukherjee}}]{pal1988molecular}%
  \BibitemOpen
  \bibfield  {author} {\bibinfo {author} {\bibfnamefont {S.}~\bibnamefont {Pal}}, \bibinfo {author} {\bibfnamefont {M.}~\bibnamefont {Rittby}}, \bibinfo {author} {\bibfnamefont {R.~J.}\ \bibnamefont {Bartlett}}, \bibinfo {author} {\bibfnamefont {D.}~\bibnamefont {Sinha}},\ and\ \bibinfo {author} {\bibfnamefont {D.}~\bibnamefont {Mukherjee}},\ }\bibfield  {title} {\bibinfo {title} {Molecular applications of multireference coupled-cluster methods using an incomplete model space: Direct calculation of excitation energies},\ }\href@noop {} {\bibfield  {journal} {\bibinfo  {journal} {The Journal of chemical physics}\ }\textbf {\bibinfo {volume} {88}},\ \bibinfo {pages} {4357} (\bibinfo {year} {1988})}\BibitemShut {NoStop}%
\bibitem [{\citenamefont {Jeziorski}\ and\ \citenamefont {Paldus}(1989)}]{jeziorski1989valence}%
  \BibitemOpen
  \bibfield  {author} {\bibinfo {author} {\bibfnamefont {B.}~\bibnamefont {Jeziorski}}\ and\ \bibinfo {author} {\bibfnamefont {J.}~\bibnamefont {Paldus}},\ }\bibfield  {title} {\bibinfo {title} {Valence universal exponential ansatz and the cluster structure of multireference configuration interaction wave function},\ }\href@noop {} {\bibfield  {journal} {\bibinfo  {journal} {The Journal of chemical physics}\ }\textbf {\bibinfo {volume} {90}},\ \bibinfo {pages} {2714} (\bibinfo {year} {1989})}\BibitemShut {NoStop}%
\bibitem [{\citenamefont {Kaldor}(1991)}]{kaldor1991fock}%
  \BibitemOpen
  \bibfield  {author} {\bibinfo {author} {\bibfnamefont {U.}~\bibnamefont {Kaldor}},\ }\bibfield  {title} {\bibinfo {title} {The fock space coupled cluster method: theory and application},\ }\href@noop {} {\bibfield  {journal} {\bibinfo  {journal} {Theoretica chimica acta}\ }\textbf {\bibinfo {volume} {80}},\ \bibinfo {pages} {427} (\bibinfo {year} {1991})}\BibitemShut {NoStop}%
\bibitem [{\citenamefont {Bernholdt}\ and\ \citenamefont {Bartlett}(1999)}]{bernholdt1999critical}%
  \BibitemOpen
  \bibfield  {author} {\bibinfo {author} {\bibfnamefont {D.~E.}\ \bibnamefont {Bernholdt}}\ and\ \bibinfo {author} {\bibfnamefont {R.~J.}\ \bibnamefont {Bartlett}},\ }\bibfield  {title} {\bibinfo {title} {A critical assessment of multireference-fock space ccsd and perturbative third-order triples approximations for photoelectron spectra and quasidegenerate potential energy surfaces},\ }in\ \href@noop {} {\emph {\bibinfo {booktitle} {Advances in Quantum Chemistry}}},\ Vol.~\bibinfo {volume} {34}\ (\bibinfo  {publisher} {Elsevier},\ \bibinfo {year} {1999})\ pp.\ \bibinfo {pages} {271--293}\BibitemShut {NoStop}%
\bibitem [{\citenamefont {Meissner}(1998)}]{meissner1998fock}%
  \BibitemOpen
  \bibfield  {author} {\bibinfo {author} {\bibfnamefont {L.}~\bibnamefont {Meissner}},\ }\bibfield  {title} {\bibinfo {title} {Fock-space coupled-cluster method in the intermediate hamiltonian formulation: Model with singles and doubles},\ }\href@noop {} {\bibfield  {journal} {\bibinfo  {journal} {The Journal of chemical physics}\ }\textbf {\bibinfo {volume} {108}},\ \bibinfo {pages} {9227} (\bibinfo {year} {1998})}\BibitemShut {NoStop}%
\bibitem [{\citenamefont {Evangelista}\ \emph {et~al.}(2007)\citenamefont {Evangelista}, \citenamefont {Allen},\ and\ \citenamefont {Schaefer}}]{evangelista2007coupling}%
  \BibitemOpen
  \bibfield  {author} {\bibinfo {author} {\bibfnamefont {F.~A.}\ \bibnamefont {Evangelista}}, \bibinfo {author} {\bibfnamefont {W.~D.}\ \bibnamefont {Allen}},\ and\ \bibinfo {author} {\bibfnamefont {H.~F.}\ \bibnamefont {Schaefer}},\ }\bibfield  {title} {\bibinfo {title} {Coupling term derivation and general implementation of state-specific multireference coupled cluster theories},\ }\href@noop {} {\bibfield  {journal} {\bibinfo  {journal} {The Journal of chemical physics}\ }\textbf {\bibinfo {volume} {127}} (\bibinfo {year} {2007})}\BibitemShut {NoStop}%
\bibitem [{\citenamefont {Lyakh}\ \emph {et~al.}(2012)\citenamefont {Lyakh}, \citenamefont {Musia{\l}}, \citenamefont {Lotrich},\ and\ \citenamefont {Bartlett}}]{mrcclyakh}%
  \BibitemOpen
  \bibfield  {author} {\bibinfo {author} {\bibfnamefont {D.~I.}\ \bibnamefont {Lyakh}}, \bibinfo {author} {\bibfnamefont {M.}~\bibnamefont {Musia{\l}}}, \bibinfo {author} {\bibfnamefont {V.~F.}\ \bibnamefont {Lotrich}},\ and\ \bibinfo {author} {\bibfnamefont {R.~J.}\ \bibnamefont {Bartlett}},\ }\bibfield  {title} {\bibinfo {title} {Multireference nature of chemistry: The coupled-cluster view},\ }\href@noop {} {\bibfield  {journal} {\bibinfo  {journal} {Chem. Rev.}\ }\textbf {\bibinfo {volume} {112}},\ \bibinfo {pages} {182} (\bibinfo {year} {2012})}\BibitemShut {NoStop}%
\bibitem [{\citenamefont {Hoffmann}(1996)}]{hoffmann1996canonical}%
  \BibitemOpen
  \bibfield  {author} {\bibinfo {author} {\bibfnamefont {M.~R.}\ \bibnamefont {Hoffmann}},\ }\bibfield  {title} {\bibinfo {title} {Canonical van vleck quasidegenerate perturbation theory with trigonometric variables},\ }\href@noop {} {\bibfield  {journal} {\bibinfo  {journal} {The Journal of Physical Chemistry}\ }\textbf {\bibinfo {volume} {100}},\ \bibinfo {pages} {6125} (\bibinfo {year} {1996})}\BibitemShut {NoStop}%
\end{thebibliography}%

\end{document}